\documentclass[10pt]{article} 
\usepackage[margin=1in]{geometry}
\usepackage{graphicx, float}
\usepackage{amsmath}
\usepackage{amssymb}
\usepackage{amsthm}
\usepackage{amsfonts}
\usepackage{enumerate}
\usepackage{tensor}
\usepackage{dsfont} 
\usepackage{cleveref}

\usepackage{caption}
\usepackage[backend=bibtex,doi=false,isbn=false,url=false,style=numeric]{biblatex}
\bibliography{ZoteroLibraryDISSERTATION} 

\newtheorem{thm}{Theorem}[section]
\newtheorem{lem}[thm]{Lemma}
\newtheorem{prop}[thm]{Proposition}
\newtheorem{cor}[thm]{Corollary}

\newtheorem{dfn}[thm]{Definition}

\newtheorem{conj}[thm]{Conjecture}

\newtheorem{fact}[thm]{Fact}
\newtheorem{Item}{Item}

\newcommand{\bthm}{\begin{thm}}
\newcommand{\ethm}{\end{thm}}
\newcommand{\blem}{\begin{lem}}
\newcommand{\elem}{\end{lem}}
\newcommand{\bcor}{\begin{cor}}
\newcommand{\ecor}{\end{cor}}
\newcommand{\bdfn}{\begin{dfn}}
\newcommand{\edfn}{\end{dfn}}
\newcommand{\bconj}{\begin{conj}}
\newcommand{\econj}{\end{conj}}
\newcommand{\bprop}{\begin{prop}}
\newcommand{\eprop}{\end{prop}}
\newcommand{\bprof}{\begin{proof}}
\newcommand{\eprof}{\end{proof}}

\newcommand{\bfig}{\begin{figure}}
\newcommand{\efig}{\end{figure}}
\newcommand{\bfact}{\begin{fact}}
\newcommand{\efact}{\end{fact}}
\newcommand{\bmin}{\begin{minipage}}
\newcommand{\emin}{\end{minipage}}
\newcommand{\bit}{\begin{Item}}
\newcommand{\eit}{\end{Item}}
\newcommand{\beq}{\begin{equation}}
\newcommand{\eeq}{\end{equation}}
\newcommand{\ba}{\begin{aligned}}
\newcommand{\ea}{\end{aligned}}

\newcommand{\be}{\begin{enumerate}}
\newcommand{\ee}{\end{enumerate}}
\newcommand{\bi}{\begin{itemize}}
\newcommand{\ei}{\end{itemize}}
\newcommand{\bfl}{\begin{flushleft}}
\newcommand{\efl}{\end{flushleft}}
\newcommand{\bc}{\begin{center}}
\newcommand{\ec}{\end{center}}
\newcommand{\bfr}{\begin{flushright}}
\newcommand{\efr}{\end{flushright}}
\newcommand{\bas}{\begin{align*}}
\newcommand{\eas}{\end{align*}}

\newcommand{\Se}{S^\epsilon}
\newcommand{\Ze}{Z^\epsilon}
\newcommand{\Ne}{N^\epsilon}
\newcommand{\We}{W^\epsilon}
\newcommand{\Qe}{\mathbb{Q}^\epsilon}

\newcommand{\Ve}{V^\epsilon}

\def\employer{\@employer}
\def\@employer#1{\noindent \par{\sc #1}}

\long\def\symbolfootnote[#1]#2{\begingroup%
\def\thefootnote{\fnsymbol{footnote}}\footnote[#1]#2\endgroup}

\title{A Gaussian Markov alternative to fractional Brownian motion for pricing financial derivatives}
\date{}

\author{
	Daniel Conus and Mackenzie Wildman}

\begin{document}

	\maketitle

	\begin{abstract}
		Replacing Black-Scholes' driving process, Brownian motion, with fractional Brownian motion allows for incorporation of a past dependency of stock prices but faces a few major downfalls, including the occurrence of arbitrage when implemented in the financial market. We present the development, testing, and implementation of a simplified alternative to using fractional Brownian motion for pricing derivatives. By relaxing the assumption of past independence of Brownian motion but retaining the Markovian property, we are developing a competing model that retains the mathematical simplicity of the standard Black-Scholes model but also has the improved accuracy of allowing for past dependence. This is achieved by replacing Black-Scholes' underlying process, Brownian motion, with a particular Gaussian Markov process, proposed by Vladimir Dobri\'{c} and Francisco Ojeda.
	\end{abstract}
	
	
\section{Introduction}\label{IntroductionChapter}
	
Under the Nobel prize-winning Black-Scholes model for pricing financial derivatives \cite{black_pricing_1973}, we assume that the underlying stock price $(S_t)_{t\in[0,\infty)}$ behaves according to the stochastic differential equation (SDE)
	\beq\label{BlackScholesSDE} dS_t = S_t(\mu\,dt + \sigma\,dW_t), \eeq
with initial condition $S(0)=S_0\in\mathbf{R^+}$ and where $\mu\in\mathbf{R}$ is the drift of the stock price, $\sigma\in(0,\infty)$ is its volatility, and $(W_t)_{t\in[0,\infty)}$ is a standard Brownian motion process. The solution to this SDE is achieved using It\^{o} calculus (see, for instance, \cite{shreve_stochastic_2004}); namely
	$$S_t = S_0 \exp\left\{\sigma W_t + \mu t - \frac{1}{2}\sigma^2 t\right\}.$$
Recall a few of the assumptions imposed by this model: the short-term interest rate $r$ is known and constant, there are no transaction costs, stock prices have constant and known volatility $\sigma$ and drift $\mu$, changes in stock price are log normally distributed, and future changes in stock price only depend on the current value and are independent of the past. The current study of Option Pricing Theory largely consists of relaxing one or more of the assumptions of the standard model and studying the new model. Incorporating a stochastic volatility into the model relaxes the assumption that the underlying stock has constant volatility as in, for example, Hull \cite{hull_pricing_1987} and Heston \cite{heston_closed-form_1993}. A Black-Scholes model that incorporates transaction costs was developed by Leland \cite{leland_option_1985}. Incorporating a jump-diffusion process instead of Brownian motion is one way to relax the Gaussian property of log returns, as first considered by Merton \cite{merton_option_1976}. Use of Brownian noise in the stock price process imposes the assumption that the log increments in stock price are independent over disjoint time intervals. One way to relax this assumption is by using fractional Brownian motion in the SDE \eqref{BlackScholesSDE} in place of Brownian motion.

Fractional Brownian motion, introduced by Mandelbrot and van Ness \cite{mandelbrot_fractional_1968}, is a Wiener process generalized to incorporate time dependence through an additional parameter, the \textit{Hurst index} $H$, which measures the intensity of long-range dependence.
	\bdfn\label{fBM}
		Fractional Brownian motion, $(Z_H(t))_{t\in[0,\infty)}$, is a real-valued centered Gaussian process, where $H\in(0,1)$, such that $Z_H(0)=0$ almost surely and
			$$\mathbb{E}[Z_H(t)Z_H(s)]=\tfrac{1}{2}\{t^{2H}+s^{2H}-|t-s|^{2H}\}.$$
	\edfn
Note that when $H=\tfrac{1}{2}$, this is equivalent to a standard Brownian motion process. For values of $H>\frac{1}{2}$, the increments of the process are positively correlated and the closer $H$ is to 1, the stronger long-memory the process exhibits. Conversely, if $H<\frac{1}{2}$, the increments of fractional Brownian motion are negatively correlated. Hu and {\O}ksendal \cite{hu_fractional_2003} and Sottinen \cite{sottinen_fractional_2001} have replaced Brownian motion with fractional Brownian motion in the Black-Scholes SDE:
	\beq\label{geofBM}dS_t = S_t(\mu\,dt + \sigma\,dZ_H(t)).\eeq
Hu and {\O}ksendal \cite{hu_fractional_2003} achieve a solution to this differential equation using Wick calculus; namely
	$$S_t = S_0 \exp\left\{\sigma Z_H(t) + \mu t - \frac{1}{2}\sigma^2 t^{2H}\right\}.$$
One motivation for incorporating past dependency of stock prices is given by an empirical study of daily returns from 1962 to 1987 \cite{scansaroli_stochastic_2012}, which shows the Hurst index 
of the S\&P 500 Index is approximately 0.61 with a 95\% confidence interval of (0.57,0.69). If the index price showed no past dependency, we would expect the Hurst index to be 0.5. (Also see arguments that log returns have long-range dependence in \cite{mandelbrot_fractals_1997} and \cite{shiryaev_essentials_1999}.) A major disadvantage, however, to this model is that it results in a non-semi-martingale stock price process. This allows for arbitrage in the financial markets 
and it fails to admit an explicit hedging strategy through the use of Wick calculus instead of It\^{o} calculus. See, for example \cite{sottinen_fractional_2001} and its references.

With these issues surrounding the use of fractional Brownian motion in mind, we introduce and implement the ``Dobri\'{c}-Ojeda process'', as originally proposed and defined by Vladimir Dobri\'{c} and Francisco Ojeda in \cite{dobric_conditional_2009}. The \textit{Dobri\'{c}-Ojeda process} is a Gaussian Markov process with similar properties to those of fractional Brownian motion, particularly dependent increments in time, and we propose this process as an alternative to fractional Brownian motion in the Black-Scholes stochastic differential equation \eqref{BlackScholesSDE}. Following \cite{dobric_conditional_2009}, we define the Dobri\'{c}-Ojeda process by first considering the fractional Gaussian field $Z=(Z_H(t))_{(t,H)\in[0,\infty)\times(0,1)}$ on a probability space $(\Omega,\mathcal{F},\mathbb{P})$ defined by the covariance
	$$\mathbb{E}\{Z_H(t)Z_{H'}(s)\}=\frac{a_{H,H'}}{2}\{|t|^{H+H'}+|s|^{H+H'}-|t-s|^{H+H'}\},$$
where
	$$a_{H,H'}=	\left\{
								\begin{array}{lr}
									-\tfrac{2}{\pi}\sqrt{\Gamma(2H+1)\sin(\pi H)}\sqrt{\Gamma(2H'+1)\sin(\pi H')}
									& \\
									\times
									\Gamma\left(-(H+H')\right)\cos\left((H'-H)\tfrac{\pi}{2}\right)\cos\left((H+H')\tfrac{\pi}{2}\right)
									& \text{for } H+H' \neq 1 \\[1ex]
									\sqrt{\Gamma(2H+1)\Gamma(3-2H)}\sin^2(\pi H) =:a_H =:a_{H'}
									& \text{for } H+H'=1,
								\end{array}
								\right.$$
where $\Gamma(t)=\int_0^{\infty}x^{t-1}e^{-x}dx$ is the usual Gamma function. Existence of this field was established in \cite{dobric_fractional_2006}. Note that when $H=H'$, $Z_H$ is a fractional Brownian motion process and when $H=H'=\tfrac{1}{2}$, $Z_H$ is a standard Brownian motion process. On this field, for the case $H+H'=1$, define the process $(M_H(t))_{t\in[0,\infty)}$, given by
	\beq\label{MHdef}
		M_H(t)
			=\mathbb{E}(Z_{H'}(t)|\mathcal{F}_t^H),
	\eeq
	where
	$$\mathcal{F}_t^H=\sigma(Z_H(r):0\leq r\leq t).$$
	As proved in \Cref{MHmartingale} below, the process $M_H$ is a martingale with respect to $(\mathcal{F}_t^H)_{t\geq0}$. This fact is stated without proof in \cite{dobric_conditional_2009}.
	The second moment of $M_H(t)$ is given by
	\beq\label{M2moment}
		\mathbb{E}[M_H^2(t)]=c_M t^{2-2H},
	\eeq
	where $c_M=\frac{a_H^2\Gamma(3/2-H)}{2H\Gamma(H+1/2)\Gamma(3-2H)}$, see \cite{dobric_conditional_2009}.
	We will also show that $M_H(t)$ is Gaussian centered with independent increments and covariance $\mathbb{E}[M_H(t)M_H(s)]=c_M \left(s \wedge t\right)^{2-2H}$ (see \Cref{Mprops}).
		
We use this process $M_H$ to capture some of the information of fractional Brownian motion by projecting a fractional Brownian motion onto the fractional Gaussian field $Z$.
	
We seek a process of the form $\Psi_H(t)M_H(t)$ that approximates fractional Brownian motion, where $\Psi_H(t)$ is some deterministic coefficient. We find such a coefficient for $M_H$ in order for the least-squares difference from $Z_H$, given by $\mathbb{E}(Z_H(t)-\Psi_H(t)M_H(t))^2$, to be minimized. Since this expectation is quadratic in $\Psi_H$, the minimizing $\Psi_H$ is given by
	$$\Psi_H(t):=\frac{\mathbb{E}(Z_H(t)M_H(t))}{\mathbb{E}M_H^2(t)}.$$
A closed form solution for $\Psi_H(t)$ is found in \cite{dobric_conditional_2009}:
	$$\Psi_H(t)=\frac{2H\Gamma(3-2H)\Gamma(H+1/2)}{a_H\Gamma(3/2-H)}t^{2H-1}:=c_{\Psi}t^{2H-1}.$$

We can finally define the Dobri\'{c}-Ojeda process $(V_H(t))_{t\in[0,\infty]}$ as
\begin{equation} \label{GMdef}
	V_H(t)=\Psi_H(t)M_H(t),
\end{equation}
where 
$$\Psi_H(t)=c_{\Psi}t^{2H-1}$$
and 
$$M_H(t)=\mathbb{E}[Z_{H'}(t)|\mathcal{F}_t^H],$$
where $H+H'=1$. 
Note that when $H=\tfrac{1}{2}$, the process $V_H(t)$ is a Brownian motion.
	
To understand how closely the Dobri\'{c}-Ojeda process $V_H$ approximates fractional Brownian motion $Z_H$, consider the difference process
		$$Y_H(t) := Z_H(t) - V_H(t).$$
As proved in \cite{dobric_conditional_2009},
		$$\mathbb{E}Y_H^2(t)=d_H^2 t^{2H}=d_H^2\mathbb{E}Z_H^2(t),$$
	for
		$$d_H^2=1-2H\frac{\Gamma(1/2+H)\Gamma(3-2H)}{\Gamma(3/2-H)}.$$

\begin{figure}[htbp]
  \centering
	\includegraphics[width=.4\textwidth,keepaspectratio=true]{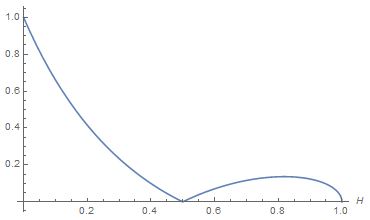}
  \caption{Graph of $d_H$.}
  \label{fig:dH}
\end{figure}

	Therefore, for $H \in (0.4,1)$, which we expect to be reasonable in most markets, $V_H$ approximates $Z_H$ with a relative $L^2$ error of at most $12\%$. We expect that $H$ is approximately 0.6 in a  typical market and rarely less than 0.4, as described and cited above. \Cref{fig:dH} shows $d_H$, which represents the relative $L^2$ error of $V_H$ from $Z_H$, as a function of $H$.
	
	
One useful property of the Dobri\'{c}-Ojeda process is that it has an It\^{o} diffusion representation and is a semi-martingale. In \Cref{GMdiffusion}, we will show that there exists a Brownain motion process $(W_t)_{t\in[0,\infty)}$ adapted to the filtration $(\mathcal{F}_t^H)_{t\in[0,\infty)}$ so that we can write
	$$dV_H(t) = Ct^{H-1/2}dW_t + (2H-1)t^{-1}V_H(t)dt,$$
where $C$ is a deterministic constant.
	
The major goal of the paper is to apply the Dobri\'{c}-Ojeda process as noise in the Black-Scholes SDE \eqref{BlackScholesSDE}:
	$$dS_t = S_t(\mu\,dt + \sigma\,dV_H(t)).$$
We emphasize that when $H=1/2$ this is equivalent to the original Black-Scholes SDE. The main advantage to the Dobri\'{c}-Ojeda process, however, is its semi-martingale property that allows for use of It\^{o} calculus. 

In order to price options, the next natural step is to describe a risk-neutral measure for this model. This does not follow directly as in the Black-Scholes model due to the $1/t$ term in the drift, as we illustrate in \Cref{NovikovFail}. This causes explosion of the expectation of the process
	$$
		\exp\left(\frac{1}{2}\int_0^t \gamma_s^2 ds \right)
	$$
at 0, where $\gamma$ is the drift correction in Girsanov's Theorem. To remedy this issue, we define a modified Dobri\'{c}-Ojeda process $(\Ve_t)_{t\in[0,\infty)}$ in which the drift is 0 until time $t=\epsilon>0$. Under the modified Dobri\'{c}-Ojeda process we achieve a risk-neutral measure $\Qe$ for the modified stock price process $(\Se_t)_{t\in[0,\infty)}$ for fixed $\epsilon>0$ using Novikov's condition \cite{revuz_continuous_1999}. In the case of a European call option, we find a price formula under this risk-neutral measure:
	$$ F_t = \Se_t\Phi\left(\sigma C\sqrt{\frac{T^{2H}-t^{2H}}{2H}}-d_1\right) - Ke^{-r(T-t)}\Phi(-d_1),$$
where $C$ is a deterministic constant and as usual in the literature, $T$ is the expiration, $K$ is the strike price, $\Phi$ is the standard normal cumulative distribution function, and
$$d_1 = \frac	{\ln\left(\frac{K}{\Se_t}\right)-r(T-t)+\frac{1}{2}\sigma^2C^2\left(\frac{T^{2H}-t^{2H}}{2H}\right)}
													{\sigma C\sqrt{\frac{T^{2H}-t^{2H}}{2H}}}.$$
Formal convergence of the measures $\Qe$ to a risk-neutral measure $\mathbb{Q}$ for $S_t$ remains an open problem.

We conclude by discussing techniques for estimating the Hurst index, $H$, and volatility, $\sigma$, using historical prices of the underlying asset, following with a comparison of historical option prices computed using Brownian motion, fractional Brownian motion, and the Dobri\'{c}-Ojeda process in the Black-Scholes SDE. We find that the model using the Dobri\'{c}-Ojeda process does, in fact, approximate the option price given using fractional Brownian motion when the parameter $H$ is similar. When using a smaller value for the Hurst index $H$, however, the Dobri\'{c}-Ojeda process appears to outperform the competing models. It is not surprising that the Dobri\'{c}-Ojeda model behaves differently from the fractional Brownian motion model for $H$ small, since for small $H$ values the two processes differ significantly (see \Cref{fig:dH}), however the improved accuracy of the Dobri\'{c}-Ojeda model for $H$ small suggests that in certain cases, stock prices do not follow a fractional Brownian motion process.

\section{The Dobri\'{c}-Ojeda process}

	In this section we will prove a few properties of the Dobri\'{c}-Ojeda process, as defined in \Cref{IntroductionChapter}. 
	
\subsection{Properties of $M_H(t)$}
	First note that the process $M_H(t)$ is Gaussian for all $t>0$ because it is the conditional expectation of a Gaussian process, $Z_H(t)$. 
	The process $M_H(t)$ also satisfies, by definition, $\mathbb{E}[M_H(t)]=0$ and, by \cite{dobric_conditional_2009}, $\mathbb{E}[M_H^2(t)]=c_M t^{2-2H}$. The following proposition is stated without proof in \cite{dobric_conditional_2009}. For the sake of completeness, we prove it here.
	
	\bprop\label{MHmartingale}
		The process $M_H(t)$ is a martingale with respect to $\mathcal{F}_t^H$. 
	\eprop
		\bprof
			Let $t>0$. By definition of $M_H(t)$, we have
			$$
							\mathbb{E}\left[\left|M_H(t)\right|\right]
				=			\mathbb{E}\left[\left|\mathbb{E}\left[Z_{1-H}(t)|\mathcal{F}_t^H\right]\right|\right] 
				\leq	\mathbb{E}\left[\mathbb{E}\left[\left|Z_{1-H}(t)\right||\mathcal{F}_t^H\right]\right] 
				=			\mathbb{E}\left[\left|Z_{1-H}(t)\right|\right]
				<			\infty,
			$$
			since $Z_{1-H}(t)$ is Gaussian.
			It remains to show that for $0\leq s<t$, $\mathbb{E}[M_H(t)|\mathcal{F}_s^H] = M_H(s)$. By the Tower Rule and by the definition of $M_H(t)$ \eqref{MHdef}, we have
			$$\ba
						\mathbb{E}[M_H(t)|\mathcal{F}_s^H]
				=&	\mathbb{E}\left[ \mathbb{E}(Z_{H'}(t)|\mathcal{F}_t^H) | \mathcal{F}_s^H \right] \\
				=&	\mathbb{E}\left[Z_{H'}(t) | \mathcal{F}_s^H \right] \\
				=&	\mathbb{E}\left[Z_{H'}(t) - Z_{H'}(s) | \mathcal{F}_s^H \right] + \mathbb{E}\left[Z_{H'}(s) | \mathcal{F}_s^H \right] \\
				=&	\mathbb{E}\left[Z_{H'}(t) - Z_{H'}(s) | \mathcal{F}_s^H \right] + M_H(s). \\
			\ea$$
			It remains to show that $\mathbb{E}\left[Z_{H'}(t) - Z_{H'}(s) | \mathcal{F}_s^H \right] = 0$.
			Fix $V\in\mathcal{F}_s^H$. Without loss of generality, let $V=\mathds{1}_{\{Z_H(u)\in B\}}$ 
			for some $u\leq s$ and where $B$ is a Borel set. Then			
			$$
						\mathbb{E}[V(Z_{H'}(t)-Z_{H'}(s))] 
				=	\mathbb{E}[\mathds{1}_{\{Z_H(u)\in B\}}Z_{H'}(t)] - \mathbb{E}[\mathds{1}_{\{Z_H(u)\in B\}}Z_{H'}(s)]. 
			$$
			First, note that for any pair of standard jointly normal random variables $X$ and $Y$ with covariance $\rho$
			and for any Borel set $B$,
			$\mathbb{E}\left[\mathds{1}_{X\in B} Y\right]=\rho\mathbb{E}\left[\mathds{1}_{X\in B} X\right]$.
			This can be easily verified by defining a third random variable, $(Y-\rho X)/\sqrt{1-\rho^2}$, which is independent of $X$.
			Moreover, for any centered jointly Gaussian random variables $X$ and $Y$ with variance $\sigma_X^2$ and $\sigma_Y^2$, respectively,
			with covariance $\rho$, and for any Borel set $B$,
			$\mathbb{E}\left[\mathds{1}_{X\in B} Y\right]=\frac{\rho}{\sigma_X^2}\mathbb{E}\left[\mathds{1}_{X\in B} X\right]$.
			Therefore, we have
			$
					\mathbb{E}[\mathds{1}_{\{Z_H(u)\in B\}}Z_{H'}(t)]
				=	\frac{a_H u}{\mathbb{E}[Z_H^2(u)]} \mathbb{E}\left[\mathds{1}_{Z_H(u)\in B}Z_H(u)\right]
			$			
			and similarly,
			$\mathbb{E}[\mathds{1}_{\{Z_H(u)\in B\}}Z_{H'}(s)]
				= \frac{a_H u}{\mathbb{E}[Z_H^2(u)]} \mathbb{E}\left[\mathds{1}_{Z_H(u)\in B}Z_H(u)\right]$.
			This shows 
			$\mathbb{E}[V(Z_{H'}(t)-Z_{H'}(s))]=0$ 
			for all random variables $V\in\mathcal{F}_s^H$
			and so 
			$\mathbb{E}\left[Z_{H'}(t) - Z_{H'}(s) | \mathcal{F}_s^H \right] = 0$.
		\eprof
		
	\bprop\label{Mprops}
		The martingale process $M_H$ has independent increments and covariance $\mathbb{E}[M_H(t)M_H(s)]=c_M \left(s \wedge t\right)^{2-2H}$.
	\eprop
	\bprof
		Assume without loss of generality that $s<t$. Then by \Cref{MHmartingale} and \eqref{M2moment} above,
		$$
			\mathbb{E}[M_H(t)M_H(s)]
			=	\mathbb{E}\left[(M_H(t)-M_H(s))M_H(s)\right]
					+ \mathbb{E}\left[(M_H(s))^2\right]
			=	c_M s^{2-2H}.
		$$
		Therefore,
		\beq\label{MHcovar}
			\mathbb{E}[M_H(t)M_H(s)] = c_M \left(s \wedge t\right)^{2-2H}.
		\eeq
		To prove independence of increments, we assume that $s<t$ and $h>0$ is small.
		Then by \eqref{MHcovar} above,
		$$
			\mathbb{E}\left[(M_H(t+h)-M_H(t))(M_H(s+h)-M_H(s))\right] = 0.
		$$
		Since the process $M_H$ is Gaussian, this suffices to show that $M_H$ has independent increments.
	\eprof
	Next we will prove that the quadratic variation of the martingale process $M_H$ from $0$ to $t$ is given by $c_M t^{2-2H}$. First we will prove the following lemma, to be used in the proof of \Cref{MQV} and later in \Cref{GMQVthm}.
	\blem\label{DeltaMapprox}
		The following approximation holds for even moments of $M_t=M_H(t)$:
		$$
			\mathbb{E}[(\Delta M_{t_i})^{2k}]
			\leq
			(2k-1)!!\left(c_M(2-2H)(t_i\wedge t_{i-1})^{1-2H}\Delta t_i\right)^k,
		$$
		where $k\geq1$ and $\Delta M_{t_i}=M_{t_i}-M_{t_{i-1}}$.
	\elem
	\bprof
		Using \eqref{M2moment} and the Mean Value Theorem,
		$$\ba
						\mathbb{E}[(\Delta M_{t_i})^2]
				=&	\mathbb{E}[M_{t_i}^2]-2\mathbb{E}[M_{t_i}M_{t_{i-1}}]+\mathbb{E}[M_{t_{i-1}}^2] \\
				=&	c_Mt_i^{2-2H}-2\mathbb{E}[(\Delta M_{t_i}+M_{t_{i-1}})M_{t_{i-1}}]+c_Mt_{i-1}^{2-2H} \\
				=&	c_Mt_i^{2-2H}-2\mathbb{E}[\Delta M_{t_i}M_{t_{i-1}}]-2\mathbb{E}[M_{t_{i-1}}^2]+c_Mt_{i-1}^{2-2H} \\
				=&	c_Mt_i^{2-2H}-2c_Mt_{i-1}^{2-2H}+c_Mt_{i-1}^{2-2H} \\
				=&	c_M(t_i^{2-2H}-t_{i-1}^{2-2H}) \\
				\leq&
						c_M(2-2H)(t_i\wedge t_{i-1})^{1-2H}\Delta t_i.
			\ea$$
		Since the process $M_t$ is Gaussian, the result follows for $k\geq1$, as required.
	\eprof
	\bprop\label{MQV}
		For $n>0$, let $t_i=\tfrac{it}{n}$, 
			$i=0,...,n$
			be a partition sequence of $[0,t]$ and $M_t=M_H(t)$ as defined in \eqref{MHdef}. 
			Then
				$$\lim_{n\rightarrow\infty}\left|\left|\sum\limits_{i=1}^n{(\Delta M_{t_i})^2} 
					- c_M t^{2-2H} \right|\right|_2 = 0$$
			and
				$$\lim_{n\rightarrow\infty}\sum\limits_{i=1}^n{(\Delta M_{t_i})^2} = c_M t^{2-2H}
					\qquad \text{a.s.}$$
			where $\Delta M_{t_i}=M_{t_i}-M_{t_{i-1}}$.
	\eprop
	\bprof
		Since the function $f(t)=t^{2-2H}$ is integrable, we have 
			$c_M(2-2H)\sum_{i=1}^n t_i^{1-2H}\Delta t \rightarrow c_M t^{2-2H}$ as $n\rightarrow\infty$ in $L^2$ and almost surely. 
		Therefore by the Triangle Inequality, it suffices to show that 
			$$\left|\left|\sum\limits_{i=1}^n{(\Delta M_{t_i})^2} 
			- c_M(2-2H) \sum_{j=1}^n t_j^{1-2H}\Delta t \right|\right|_2 \rightarrow0$$
		as $n\rightarrow\infty$.
		Using the independent increments of $M_H$ as proved in \Cref{Mprops}, we have
		$$\ba
			&		\left|\left|\sum\limits_{i=1}^n{(\Delta M_{t_i})^2} 
					- c_M(2-2H) \sum_{j=1}^n t_j^{1-2H}\Delta t \right|\right|_2 \\
			=&	\mathbb{E}\left[\left(\sum\limits_{i=1}^n{(\Delta M_{t_i})^2} 
					- c_M(2-2H) \sum_{j=1}^n t_j^{1-2H}\Delta t \right)^2 \right] \\
			=&	\sum\limits_{i=1}^n\sum\limits_{j=1}^n\mathbb{E}\left[(\Delta M_{t_i})^2\right]\mathbb{E}\left[(\Delta M_{t_j})^2\right]
					- 2c_M(2-2H) \sum\limits_{i=1}^n\sum_{j=1}^n \mathbb{E}\left[(\Delta M_{t_i})^2\right] t_j^{1-2H}\Delta t \\
					&+ c_M^2 (2-2H)^2 \sum_{i=1}^n\sum_{j=1}^n t_i^{1-2H} t_j^{1-2H} (\Delta t)^2. \\
		\ea$$
		By \Cref{DeltaMapprox}, this is bounded above by 0 both in the case $H<1/2$ and $H\geq1/2$. Borel Cantelli implies almost-sure convergence.
	\eprof
	
\subsection{Properties of $V_H(t)$}
	Next, we show that the Dobri\'{c}-Ojeda process has an It\^{o} diffusion representation.
	\bprop\label{GMdiffusion}
		There exists a Brownian motion process $(W_t)_{t\in[0,\infty)}$ adapted to the filtration $(\mathcal{F}_t^H)_{t\in[0,\infty)}$ such that the Dobri\'{c}-Ojeda process $(V_H(t))_{t\in[0,\infty)}$ as defined in \eqref{GMdef}, is an It\^{o} diffusion process that satisfies the stochastic differential equation
			$$dV_H(t) = Ct^{H-1/2}dW_t + (2H-1)t^{-1}V_H(t)dt,$$
		where $C=c_\Psi \sqrt{c_M (2-2H)}$. 
	\eprop
	\bprof
		By \Cref{MQV}, the quadratic variation of $M_H$ is given by $[M_H,M_H]_t=c_M t^{2-2H}$. Therefore by the Representation Theorem for Martingales (see \cite[Thm~4.2]{karatzas_brownian_1998}), there exists a Brownian motion process $W_t$ adapted to the filtration $(\mathcal{F}_t^H)_{t\in[0,\infty)}$ for which $dM_H(t)=\sqrt{c_M(2-2H)}t^{1/2-H}dW_t$.
		 Therefore,
		$$
		\ba
			dV_H(t)
			&=	d(\Psi_H(t)M_H(t)) \\
			&=	\Psi_H(t)dM_H(t) + M_H(t)d\Psi_H(t) \\
			&=	\Psi_H(t)\sqrt{c_M(2-2H)}t^{1/2-H}dW_t + (\Psi_H(t)^{-1}V_H(t))d(c_{\Psi}t^{2H-1}) \\
			&=	c_{\Psi}\sqrt{c_M(2-2H)}t^{2H-1}t^{1/2-H}dW_t + c_{\Psi}^{-1}t^{-2H+1}V_H(t)c_{\Psi}(2H-1)t^{2H-2}dt \\
			&=	c_{\Psi}\sqrt{c_M(2-2H)}t^{H-1/2}dW_t + (2H-1)t^{-1}V_H(t)dt.
		\ea
		$$
		Notice that this equation is well-defined since $V_H(t)$ is of the order $t^H$, despite the factor $\frac{1}{t}$ in the drift.
		Also note that we can write this diffusion as
		$$
			dV_H(t) = c_{\Psi}\sqrt{c_M(2-2H)}t^{H-1/2}dW_t + c_\Psi(2H-1)t^{2H-2}M_H(t)dt,
		$$
		using the definition of $V_H(t)$ \eqref{GMdef}.
	\eprof
	The martingale part of this representation has a similar form to the Riemann-Liouville fractional integral
		$Z_H(t)=\frac{1}{\Gamma(H+1/2)} \int_0^t (t-s)^{H-1/2}dW_s$ (see \cite{biagini_stochastic_2008}), 
		but is non-anticipating and therefore It\^{o}-integrable while the fractional integral is not.
		 We consider that the drift term of the diffusion somehow compensates for this difference and works to
	imitate fractional Brownian motion while remaining a semi-martingale process.
	
	A closed-form equation for the quadratic variation of the Dobri\'{c}-Ojeda process immediately follows:
	\bcor\label{GMQV}
		The quadratic variation of $(V_H(t))_{t\in[0,\infty)}$ is given by
			$$ [V_H,V_H]_t = \frac{C^2}{2H}t^{2H}, $$
		where $C=c_\Psi \sqrt{c_M (2-2H)}$, as above.
	\ecor
\section{Option pricing with the Dobri\'{c}-Ojeda process}\label{sec:DOandoptionpricing}

	We replace Brownian motion with the Dobri\'{c}-Ojeda process in the Black-Scholes stochastic differential equation:
		$$
			dS_t=S_t(\mu dt + \sigma dV_t ).
		$$
	To simplify notation, we drop the subscript $H$ from $V_H(t)$, $M_H(t)$, and $\mathcal{F}_t^H$. Note that when $H=1/2$, we have a geometric Brownian motion process, so without loss of generality, we assume $H\neq 1/2$. Using It\^{o} calculus, we can solve for $S_t$ explicitly:
	Let $Y_t = \ln S_t$. Then we have
			$$
				\ba
							dY_t
					&=	\frac{dS_t}{S_t} - \frac{1}{2}\frac{(dS_t)^2}{(S_t)^2} \\
					&=	\mu dt + \sigma dV_t - \frac{1}{2}\sigma^2 d[V,V]_t,
				\ea
			$$
	and thus by \Cref{GMQV},
			$$
				\ba
							Y_t 
					&=	Y_0 + \mu t + \sigma V_t - \frac{1}{2} \sigma^2 [V,V]_t \\
					&=	Y_0 + \mu t + \sigma V_t - \frac{1}{2} \sigma^2 \frac{C^2}{2H}t^{2H},
				\ea
			$$
	 which implies
		\beq\label{geoGMnoe}
			S_t = S_0\exp\left\{\mu t + \sigma V_t - \frac{C^2\sigma^2}{4H} t^{2H} \right\}.
		\eeq
	As in the original model, we define $(B_t)_{t\in[0,\infty)}$ to be the bond price process with risk-free deterministic constant interest rate $r>0$, i.e. $dB_t=rB_t\,dt$, or $B_t=e^{rt}$ for all $t\geq0$.
	
\subsection{Risk-neutral measure}\label{sec:riskneutralmeasure}
	
	The next natural step towards a comprehensive model for derivative pricing is to establish the existence of a risk-neutral measure, i.e. a measure equivalent to our original measure $\mathbb{P}$ under which the discounted stock price process,
	$$dZ_t = Z_t(\sigma dV_t + (\mu-r)dt),$$
	is a martingale.
	By \Cref{GMdiffusion}, we have
	$$dZ_t = \sigma C t^{H-1/2}Z_t \left( dW_t + \gamma_t dt\right),$$
	where
	\beq\label{gammadefFail} \gamma_t = \frac{2H-1}{C}t^{-1/2-H}V_t + \frac{\mu-r}{\sigma C}t^{1/2-H}. \eeq
	The standard technique to achieve a risk-neutral measure $\mathds{Q}$ is to invoke Girsanov's Theorem by showing that $\gamma_t$ satisfies Novikov's Condition or Kazamaki's Condition (see \cite[Ch~8, \S1]{revuz_continuous_1999}).
	To date, this remains an open problem as the usual techniques fail to work in this case. For example, we will show that Novikov's Condition fails to be satisfied in the following proposition.
	
	\bprop\label{NovikovFail}
		For $0\leq t\leq T$ and for $\gamma_t$ as defined in \eqref{gammadefFail}, we have
		\beq\label{NovikovFaileq}
			\mathbb{E}\left[\exp\left(\frac{1}{2}\int_0^t \gamma_s^2 ds \right)\right]=\infty.
		\eeq
	\eprop
		\bprof
			We can write
			$$ \gamma_s^2 = A^2 s^{-1-2H}V_s^2 + 2AB s^{-2H}V_s + B^2 s^{1-2H}, $$
			where $A$ and $B$ are deterministic and constant.
			Therefore, we have
			$$\ba
				&			\mathbb{E}\left[\exp\left(\frac{1}{2}\int_0^t \gamma_s^2 ds \right)\right]
				\geq	\exp\left(\mathbb{E}\left[\frac{1}{2}\int_0^t \left(A^2 s^{-1-2H}V_s^2 + 2AB s^{-2H}V_s + B^2 s^{1-2H}\right) ds \right]\right) \\
				&=		\exp\left(\frac{A^2c_M}{2}\int_0^t  s^{-1} ds \right)
							\exp\left(\frac{B^2}{2(2-2H)} t^{2-2H} \right)
				=		\infty,
			\ea$$
			by Jensen's Inequality and properties of $V_t$.
		\eprof
	
	The determination of a risk-neutral probability measure without using Girsanov's Theorem remains an open problem. In the meantime, to resolve this issue and find a risk-neutral measure, 
	we propose to replace $V_t$ with $\Ve_t$, defined to be slightly altered from the diffusion process given in \Cref{GMdiffusion}. Since the issue lies in the $1/t$ term of the drift, we simply ``turn off'' 
	the drift until some time $\epsilon>0$. We can proceed with the standard techniques, as in \cite{shreve_stochastic_2004}, using the modified Dobri\'{c}-Ojeda process $\Ve_t$ in the stock price SDE.
	\begin{dfn}\label{GMediffusion}
		Let $\epsilon>0$. Define the Modified Dobri\'{c}-Ojeda process, $(\Ve_t)_{t\in[0,\infty)}$, by
		\beq\label{eqn:GMediffusion}
			d\Ve_t = Ct^{H-1/2}dW_t + c_\Psi(2H-1)t^{2H-2}M_t \mathds{1}_{[\epsilon,\infty)}(t)dt,
		\eeq
		where $C=c_\Psi \sqrt{c_M (2-2H)}$.
	\end{dfn}
	The drift part of $V_t$ which causes \eqref{NovikovFaileq} to explode at time $t=0$, is 0 until it ``turns on'' at time $t=\epsilon$ for any \textit{admissible} $\epsilon>0$, as we will see in \Cref{Novikov}.
	We will proceed towards derivative pricing using the model driven by $\Ve_t$ and define an option price. We begin by proving a few properties about $\Ve_t$.
	First, we show that both integrals in \eqref{eqn:GMediffusion} are well-defined.
			Using It\^{o} Isometry,
			\beq\label{tlessthane2moment}
					\mathbb{E}\left[\left(C\int_0^t s^{H-1/2}\,dW_s\right)^2\right]
				=	C^2\int_0^t s^{2H-1}\,ds
				=	\frac{C^2}{2H}t^{2H}
				<	\infty.
			\eeq
			For $t\leq\epsilon$, the second integral is 0.
			To show that the second integral is well-defined for $t>\epsilon$, 
			using \Cref{Mprops}, we have
			\beq\ba\label{tgreaterthane2moment}
				&		\mathbb{E}\left[\left(c_\Psi(2H-1)\int_0^t s^{2H-2}M_s \mathds{1}_{[\epsilon,\infty)}(s)\,ds\right)^2\right] \\
				=&	c_\Psi^2(2H-1)^2\int_\epsilon^t\int_\epsilon^t s_1^{2H-2}s_2^{2H-2}\mathbb{E}\left[M_{s_1}M_{s_2}\right]\,ds_2\,ds_1 \\
				=&	c_\Psi^2(2H-1)^2\int_\epsilon^t\int_\epsilon^t s_1^{2H-2}s_2^{2H-2}c_M(s_1\wedge s_2)^{2-2H}\,ds_2\,ds_1 \\
				=&	2c_Mc_\Psi^2(2H-1)^2 \left(\frac{1}{2H}(t^{2H}-\epsilon^{2H})-\frac{\epsilon}{2H-1}(t^{2H-1}-\epsilon^{2H-1})\right)
				<	\infty.
			\ea\eeq
		This suffices to show that $\Ve_t$, as in \eqref{eqn:GMediffusion} is well-defined.
	\bprop\label{GMemoments}
		The modified Dobri\'{c}-Ojeda process $(\Ve_t)_{t\in[0,\infty)}$ satisfies, for all $\epsilon>0$,
		\be
			\item $\mathbb{E}[\Ve_t]=0$ for all $t>0$ \text{ and}
			\item $\mathbb{E}[(\Ve_t)^2]
						=
						\left\{
						\begin{array}{ll}
							\frac{C^2t^{2H}}{2H}
								& \text{ if } t\leq\epsilon \\[1.5ex]
								  \frac{C^2}{2H}t^{2H}
										+ 2C^2(2H-1)\frac{1}{2H}(t^{2H}-\epsilon^{2H}) & \\
									+ 2c_Mc_\Psi^2(2H-1)^2 \left(\frac{1}{2H}(t^{2H}-\epsilon^{2H}) \right. & \\
									\left.-\frac{\epsilon}{2H-1}(t^{2H-1}-\epsilon^{2H-1})\right)
								& \text{ if } t>\epsilon.
						\end{array}
						\right.$
		\ee
	\eprop
		\bprof 
			\be
				\item 
				For $t\leq\epsilon$, by \Cref{GMediffusion}, we have
							$$
								\mathbb{E}[\Ve_t]
								=	\mathbb{E}\left[C\int_0^t s^{H-1/2}\,dW_s\right]
								= 0
							$$
							since it's the expectation of a square-integrable It\^{o} integral.
							For $t>\epsilon$, because the process $(M_t)$ is a martingale and thus has zero expectation, we have
							$$\ba
								\mathbb{E}[\Ve_t]
								=&	\mathbb{E}\left[C\int_0^t s^{H-1/2}\,dW_s + c_\Psi(2H-1)\int_0^t s^{2H-2}M_s \mathds{1}_{[\epsilon,\infty)}(s)\,ds\right] \\
								=&	\mathbb{E}\left[C\int_0^t s^{H-1/2}\,dW_s\right] 
										+ \mathbb{E}\left[c_\Psi(2H-1)\int_0^t s^{2H-2}M_s \mathds{1}_{[\epsilon,\infty)}(s)\,ds\right] \\
								=&	c_\Psi(2H-1)\int_0^t s^{2H-2}\mathbb{E}[M_s] \mathds{1}_{[\epsilon,\infty)}(s)\,ds
								=	0.
							\ea$$
				\item	
				For $t\leq\epsilon$, we have
							$$
								\mathbb{E}\left[\left(\Ve_t\right)^2\right]
								=	\frac{C^2}{2H}t^{2H}
							$$
							as in \eqref{tlessthane2moment} above.
							For $t>\epsilon$, as in \eqref{tgreaterthane2moment} above, we have
							$$\ba
								&		\mathbb{E}\left[\left(\Ve_t\right)^2\right] \\
								=&	\mathbb{E}\left[\left(C\int_0^t s^{H-1/2}\,dW_s 
										+ c_\Psi(2H-1)\int_0^t s^{2H-2}M_s \mathds{1}_{[\epsilon,\infty)}(s)\,ds\right)^2\right] \\
								=&	\frac{C^2}{2H}t^{2H}
										+ 2Cc_\Psi(2H-1)\mathbb{E}\left[\int_\epsilon^t\int_0^t s_1^{H-1/2}s_2^{2H-2}M_{s_2}\,dW_{s_1}\,ds_2\right] \\
										&+ 2c_Mc_\Psi^2(2H-1)^2 \left(\frac{1}{2H}(t^{2H}-\epsilon^{2H})-\frac{\epsilon}{2H-1}(t^{2H-1}-\epsilon^{2H-1})\right) \\
								=&	\frac{C^2}{2H}t^{2H}
										+ 2C^2(2H-1)\frac{1}{2H}(t^{2H}-\epsilon^{2H}) \\
										&+ 2c_Mc_\Psi^2(2H-1)^2 \left(\frac{1}{2H}(t^{2H}-\epsilon^{2H})-\frac{\epsilon}{2H-1}(t^{2H-1}-\epsilon^{2H-1})\right).
							\ea$$
							Note that the middle term can be computed using the same Martingale representation as in the proof of \Cref{GMdiffusion}:
							$$\ba
								&		\mathbb{E}\left[\int_0^t\int_\epsilon^t s_1^{H-1/2}s_2^{2H-2}M_{s_2}\,dW_{s_1}\,ds_2\right] \\
								=&	\int_\epsilon^ts_2^{2H-2}\mathbb{E}\left[M_{s_2}\int_0^t s_1^{H-1/2}\,dW_{s_1}\right]\,ds_2 \\
								=&	\sqrt{c_M(2-2H)}\int_\epsilon^ts_2^{2H-2}\mathbb{E}\left[\int_0^{s_2}u^{1/2-H}\,dW_u\int_0^t s_1^{H-1/2}\,dW_{s_1}\right]\,ds_2 \\
								=&	\sqrt{c_M(2-2H)}\int_\epsilon^ts_2^{2H-2}\int_0^{s_2\wedge t}\,du\,ds_2 \\
								=&	\frac{\sqrt{c_M(2-2H)}}{2H}(t^{2H}-\epsilon^{2H}). \\
							\ea$$
			\ee
			This concludes the proof of \Cref{GMemoments}.
		\eprof
	The quadratic variation of the modified Dobri\'{c}-Ojeda process follows immediately from \Cref{GMediffusion}.
	\bprop\label{GMeQV}
		The quadratic variation of $(\Ve_t)_{t\in[0,\infty)}$ is given by
			$$ [\Ve,\Ve]_t = \frac{C^2}{2H}t^{2H}, $$
		where $C=c_\Psi \sqrt{c_M (2-2H)}$, as above.
	\eprop
	The modified Dobri\'{c}-Ojeda process has the same quadratic variation as the original Dobri\'{c}-Ojeda process because while the drift component has been modified, only the martingale part contributes to the quadratic variation.
	\bprop\label{GMeconverges}
		For $H\in(0,1)$ fixed, the process $(\Ve_t)_{t\in[0,\infty)}$ as defined in \Cref{GMediffusion} converges uniformly in $t$ both in $L^2(\Omega)$ and almost surely to the original Dobri\'{c}-Ojeda process $(V_t)_{t\in[0,\infty)}$ as $\epsilon\rightarrow0$.
	\eprop
		\bprof
			For $\epsilon>0$, define the process $(\Ne_t)_{t\in[0,\infty)}$ by
			$$
				\Ne_t = V_t - \Ve_t
			$$
			for all $t\geq 0$. Then by \Cref{GMdiffusion}, \Cref{GMediffusion}, and the original definition of the
			Dobri\'{c}-Ojeda process \eqref{GMdef},
			$$\ba
						d\Ne_t 	
				&= 	dV_t - d\Ve_t \\
				&=	(2H-1)\left(t^{-1}V_t -c_\Psi t^{2H-2}M_t \mathds{1}_{[\epsilon,\infty)}(t)\right)dt \\
				&=	(2H-1)t^{-1}\left(V_t - V_t \mathds{1}_{[\epsilon,\infty)}(t)\right)dt \\
				&=	\left\{\begin{array}{ll}
							(2H-1)t^{-1}V_t dt		& \text{ if } t<\epsilon \\
							0											& \text{ if } t\geq\epsilon.
						\end{array}
						\right.
			\ea$$
			When $t<\epsilon$,
			$$
				\Ne_t = (2H-1)\int_0^t s^{-1}V_s ds.
			$$
			When $t\geq\epsilon$, $d\Ne_t=0$.
			Therefore, $\Ne_t$ is constant for $t\geq\epsilon$, with 
			$$\Ne_t=\Ne_\epsilon=(2H-1)\int_0^\epsilon s^{-1}V_s ds.$$
			Then by the Minkowski and Cauchy-Schwarz inequalities,
			$$\ba
							\mathbb{E}\left[\sup_{0\leq t<\infty}\left(\Ne_t\right)^2\right]
				=&		\mathbb{E}\left[\sup_{0\leq t\leq\epsilon}\left|(2H-1)\int_0^t s^{-1}V_s \,ds\right|^2\right] \\
				\leq&	(2H-1)^2\mathbb{E}\left[\sup_{0\leq t\leq\epsilon}\left(\int_0^t |s^{-1}V_s| \,ds\right)^2\right] \\
				=&		(2H-1)^2\int_0^\epsilon\int_0^\epsilon s^{-1}r^{-1}\mathbb{E}\left[|V_sV_r|\right] \,ds\,dr \\
				\leq&	(2H-1)^2\int_0^\epsilon\int_0^\epsilon s^{-1}r^{-1}||V_s||_2||V_r||_2 \,ds\,dr \\
				=&		c_\Psi^2 c_M(2H-1)^2 \left(\int_0^\epsilon s^{H-1} \,ds\right)^2 \\
				=&		\frac{c_\Psi^2 c_M(2H-1)^2}{H^2}\epsilon^{2H} 
				\rightarrow 0
			\ea$$
			as $\epsilon\rightarrow0$. $L^2$ convergence follows directly.
			Almost-sure convergence is straight-forward using the Dominated Convergence Theorem:
			$$\ba
						\lim_{\epsilon\rightarrow0}\Ve_t
				&=	\lim_{\epsilon\rightarrow0}
							\left(C\int_0^t s^{H-1/2}\,dW_s + c_\Psi(2H-1)\int_0^t s^{2H-2}M_s \mathds{1}_{[\epsilon,\infty)}(s)\,ds\right) \\
				&=	C\int_0^t s^{H-1/2}\,dW_s + \lim_{\epsilon\rightarrow0}c_\Psi(2H-1)\int_0^t s^{2H-2}M_s \mathds{1}_{[\epsilon,\infty)}(s)\,ds \\
				&=	C\int_0^t s^{H-1/2}\,dW_s + c_\Psi(2H-1)\int_0^t s^{2H-2}M_s \lim_{\epsilon\rightarrow0}\mathds{1}_{[\epsilon,\infty)}(s)\,ds \\
				&=	C\int_0^t s^{H-1/2}\,dW_s + c_\Psi(2H-1)\int_0^t s^{2H-2}M_s\,ds, \\
				&=	V_t,
			\ea$$
			where the indicator function is simply bounded by 1.
		\eprof
	With these properties of the Modified Dobri\'{c}-Ojeda process in mind, we proceed towards pricing options by next defining a modified stock price process, $\Se_t$:
		\beq\label{SteDEF}
			d\Se_t = \Se_t(\sigma d\Ve_t + \mu dt ).
		\eeq
	We will assume that the underlying stock price process follows $(\Se_t)_{t\in[0,\infty)}$, for some small $\epsilon>0$.
	By \Cref{GMediffusion}, we can use It\^{o} Calculus to solve: Let $Y_t = \ln \Se_t$. Then we have
			$$
				\ba
							dY_t
					&=	\frac{d\Se_t}{\Se_t} - \frac{1}{2}\frac{(d\Se_t)^2}{(\Se_t)^2} \\
					&=	\mu dt + \sigma d\Ve_t - \frac{1}{2}\sigma^2 d[\Ve,\Ve]_t,
				\ea
			$$
	and thus by \Cref{GMeQV},
			$$
				\ba
							Y_t 
					&=	Y_0 + \mu t + \sigma \Ve_t - \frac{1}{2} \sigma^2 [\Ve,\Ve]_t \\
					&=	Y_0 + \mu t + \sigma \Ve_t - \frac{1}{2} \sigma^2 \frac{C^2}{2H}t^{2H},
				\ea
			$$
	 which implies
		\beq\label{geoGM}
			\Se_t = S_0\exp\left\{\mu t + \sigma \Ve_t - \frac{C^2\sigma^2}{4H} t^{2H} \right\}.
		\eeq
	Since $\Ve_t$ converges to $V_t$ almost surely, convergence of $\Se_t$ 
	to $S_t$, as in \eqref{geoGMnoe}, immediately follows.
		
	Define
		$$
			\ba
						\Ze_t
				&:= B_t^{-1}\Se_t \\
				&= 	S_0\exp\left\{(\mu-r) t + \sigma \Ve_t - \frac{C^2\sigma^2}{4H} t^{2H} \right\},
			\ea
		$$
	where $B_t=e^{rt}$ is the bond price process.
		
	Then by It\^{o}'s Lemma and by \Cref{GMediffusion}, we have
		$$
			\ba
						d\Ze_t 
				=&	\Ze_t\left(\sigma d\Ve_t + (\mu-r)dt\right) \\
				=&	\Ze_t\bigg(\sigma \Big(Ct^{H-1/2}dW_t + c_\Psi(2H-1)t^{2H-2}M_t \mathds{1}_{[\epsilon,\infty)}(t)dt\Big) + (\mu-r)dt\bigg) \\
				=&	\Ze_t\bigg(\sigma Ct^{H-1/2}dW_t + \Big(\mu-r + \sigma c_\Psi(2H-1)t^{2H-2}M_t \mathds{1}_{[\epsilon,\infty)}(t)\Big) dt\bigg) \\
				=&	\sigma Ct^{H-1/2}\Ze_t\left(dW_t + 
							\left(\frac{\mu-r}{\sigma C}t^{1/2-H} + \frac{c_\Psi(2H-1)M_t \mathds{1}_{[\epsilon,\infty)}(t)}{C}t^{H-3/2}\right) dt\right). \\
			\ea
		$$
		
	Let
	\beq\label{gammadef}
		\gamma_t = At^{1/2-H} + Bt^{H-3/2}M_t \mathds{1}_{[\epsilon,\infty)}(t),
	\eeq
	where
	\beq
		A = \frac{\mu-r}{\sigma C} \qquad \text{ and } \qquad B = \frac{c_\Psi(2H-1)}{C}.
	\eeq
	
	In order to employ Girsanov's Theorem, we first verify Novikov's Condition (see \cite{karatzas_brownian_1998}), which will be satisfied for restricted values of $\epsilon$.
	This restriction is discussed following the proof.
	
\begin{figure}[htbp]
  \centering
	\includegraphics[width=.4\textwidth,keepaspectratio=true]{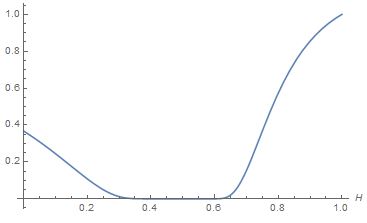}
  \caption{Graph of $\delta(H)$.}
  \label{fig:deltaHfigure}
\end{figure}

	\bprop\label{Novikov}
		For $\gamma_t$ as defined in \eqref{gammadef} and for $\epsilon>e^{\frac{-1}{2B^2c_M}}T$,
		$$
			\mathbb{E}\left[\exp\left(\frac{1}{2}\int_0^t \gamma_s^2 \,ds\right)\right] < \infty,
		$$
		for all $0<t\leq T$.
	\eprop
	\bprof
		By the Cauchy-Schwarz inequality, we have
		$$\ba
			&		\mathbb{E}\left[\exp\left(\frac{1}{2}\int_0^t \gamma_s^2 \,ds\right)\right] \\
			=&	\mathbb{E}\left[\exp\left(\frac{1}{2}\int_0^t (As^{1/2-H} + Bs^{H-3/2}M_s \mathds{1}_{[\epsilon,\infty)}(s))^2 \,ds\right)\right] \\
			=&	e^{\frac{A^2 t^{2-2H}}{2(2-2H)}}
													\mathbb{E}\left[\exp\left(AB\int_0^t s^{-1}M_s\mathds{1}_{[\epsilon,\infty)}(s) \,ds\right)
													\exp\left(\frac{1}{2}B^2\int_0^t s^{2H-3}M_s^2 \mathds{1}_{[\epsilon,\infty)}(s) \,ds\right)\right] \\
			\leq&	e^{\frac{A^2 t^{2-2H}}{2(2-2H)}}
											\left(\mathbb{E}\left[\exp\left(2AB\int_\epsilon^t s^{-1}M_s\,ds\right)\right]\right)^{1/2}
											\left(\mathbb{E}\left[\exp\left(B^2\int_\epsilon^t s^{2H-3}M_s^2\,ds\right)\right]\right)^{1/2}. \\
		\ea$$
		Note that we can use the moment generating function of the Gaussian random variable 
		$\int_\epsilon^t s^{-1}M_s\,ds$ to show that the first term is finite.
		To show that the last term is finite, first note that for $k\geq1$ and $(B_t)_{t\in[0,\infty)}$ a Brownian motion process,
		\beq\ba\label{banditbarksatUPS}
					\int_{c_M\epsilon^{2-2H}}^{c_Mt^{2-2H}} r^{-2} \mathbb{E}\left[B_r^{2k}\right]^{1/k} \,dr
			=&	\int_{c_M\epsilon^{2-2H}}^{c_Mt^{2-2H}} r^{-2} \left(\frac{2^k\Gamma(k+1/2)}{\sqrt{\pi}} r^k\right)^{1/k} \,dr \\
			=&	\frac{2\Gamma(k+1/2)^{1/k}(2-2H)}{\pi^{1/2k}} \ln\left(\frac{t}{\epsilon}\right). \\
		\ea\eeq
		By the Time-Change for Martingales (see \cite[p~174,Thm~4.6]{karatzas_brownian_1998}) and \Cref{MQV}, we can write $M_t$ as $B_{<M>_t}=B_{c_Mt^{2-2H}}$ for any $t\geq0$, where $(B_t)_{t\in[0,\infty)}$ is a Brownian motion process adapted to $(\mathcal{F}_t)_{t\in[0,\infty)}$. Note that the notation $B_t$ used in this proof is unrelated to the bond price process of the same name, used outside of \Cref{sec:riskneutralmeasure}.
		Using the Taylor expansion of $f(x)=e^x$ along with this time change, we have
		$$\ba
					\mathbb{E}\left[\exp\left(B^2\int_\epsilon^t s^{2H-3}M_s^2\,ds\right)\right]
			=&	\mathbb{E}\left[\exp\left(B^2\int_\epsilon^t s^{2H-3}B_{c_Ms^{2-2H}}^2\,ds\right)\right] \\
			=&	\mathbb{E}\left[\exp\left(B^2c_M\int_\epsilon^t s^{2H-3}B_{s^{2-2H}}^2\,ds\right)\right] \\
			=&	\mathbb{E}\left[\sum_{k=0}^\infty\frac{1}{k!}\left(\frac{B^2c_M}{2-2H}\int_{\epsilon^{2-2H}}^{t^{2-2H}} r^{-2} B_r^2 \,dr\right)^k\right]. \\
		\ea$$
		Then using Cauchy-Schwarz inequality, we have
		$$\ba
			&		\mathbb{E}\left[\sum_{k=0}^\infty\frac{1}{k!}\left(\frac{B^2c_M}{2-2H}\int_{\epsilon^{2-2H}}^{t^{2-2H}} r^{-2} B_r^2 \,dr\right)^k\right] \\
			=&	\sum_{k=0}^\infty\frac{(B^2c_M)^k}{(2-2H)^k k!}\,
						\int_{\epsilon^{2-2H}}^{t^{2-2H}}\ldots\int_{\epsilon^{2-2H}}^{t^{2-2H}} 
								r_1^{-2}\ldots r_k^{-2} \mathbb{E}\left[B_{r_1}^2\ldots B_{r_k}^2\right] \,dr_1\ldots\,dr_k \\
			\leq& 1 +	\sum_{k=1}^\infty\frac{(B^2c_M)^k}{(2-2H)^k k!}\,
						\int_{\epsilon^{2-2H}}^{t^{2-2H}}\ldots\int_{\epsilon^{2-2H}}^{t^{2-2H}} r_1^{-2}\ldots r_k^{-2} \\
						&\times\mathbb{E}\left[B_{r_1}^{2k}\right]^{1/k} \ldots \mathbb{E}\left[B_{r_k}^{2k}\right]^{1/k} \,dr_1\ldots\,dr_k \\
			=&	1 + \sum_{k=1}^\infty\frac{(B^2c_M)^k}{(2-2H)^k k!}
						\left( \int_{\epsilon^{2-2H}}^{t^{2-2H}} r^{-2} \mathbb{E}\left[B_r^{2k}\right]^{1/k} \,dr \right)^k.
		\ea$$
		Finally, by \eqref{banditbarksatUPS}, we have
		$$\ba
			&		1 + \sum_{k=1}^\infty\frac{(B^2c_M)^k}{(2-2H)^k k!}
						\left( \int_{\epsilon^{2-2H}}^{t^{2-2H}} r^{-2} \mathbb{E}\left[B_r^{2k}\right]^{1/k} \,dr \right)^k \\
			=&	1 + \frac{1}{\sqrt{\pi}}\sum_{k=1}^\infty\frac{(2B^2c_M)^k\Gamma(k+1/2)}{k!} \left( \ln\left(\frac{t}{\epsilon}\right) \right)^k \\
			\leq&	1 + \frac{1}{\sqrt{\pi}}\sum_{k=1}^\infty\frac{(2B^2c_M)^k\Gamma(k+1)}{k!} \left( \ln\left(\frac{t}{\epsilon}\right) \right)^k \\
			=&	1 + \frac{1}{\sqrt{\pi}}\sum_{k=1}^\infty \left( 2B^2c_M\ln\left(\frac{t}{\epsilon}\right) \right)^k. \\
		\ea$$
		This series converges when
		$$\ba
			\left|2B^2c_M\ln\left(\frac{t}{\epsilon}\right)\right|<1,
		\ea$$
		or when
		$$
			te^{\frac{-1}{2B^2c_M}} < \epsilon < te^{\frac{1}{2B^2c_M}},
		$$
		in which case $\mathbb{E}\left[\exp\left(\frac{1}{2}\int_0^t \gamma_s^2 \,ds\right)\right]<\infty$.
		\eprof
		The right-hand inequality is irrelevant since $te^{\frac{1}{2B^2c_M}}>t$ and we intend for $\epsilon$ to be small.
		The left-hand inequality, $\epsilon > te^{\frac{-1}{2B^2c_M}}$, has more important implications.
	To further consider this restriction on $\epsilon$, set
	$$\delta(H)=e^{\frac{-1}{2B^2c_M}}.$$
	A graph of $H\rightarrow\delta(H)$ is illustrated in \Cref{fig:deltaHfigure}.
	We do expect \Cref{Novikov} to be satisfied for \textit{any} $\epsilon>0$ since intuitively, the Brownian motion process $B_t$ behaves like
	$\sqrt{t}$ and the second term can be approximated (non-rigorously) by
		$$
			\exp\left(B^2c_M\int_\epsilon^t s^{-1} \,ds\right) < \infty,
		$$
	however a rigorous proof of the theorem for any $\epsilon>0$ remains a work in progress.	
		
	By Girsanov's Theorem (see \cite[Ch~8, Thm~1.4]{revuz_continuous_1999}),
	there exists a measure $\Qe$ equivalent to our original measure $\mathbb{P}$ such that
		$$\ba
						d\We_t
				&= 	dW_t + \gamma_t dt \\
				&=	dW_t + \left( At^{1/2-H} + Bt^{H-3/2}M_t \mathds{1}_{[\epsilon,\infty)}(t) \right) dt
		\ea$$
	is a Brownian motion process under $\Qe$.
	Therefore,
		$$\ba
					d\Ze_t
			&=	\sigma C t^{H-1/2}\Ze_t \left(dW_t + \left( At^{1/2-H} + Bt^{H-3/2}M_t \mathds{1}_{[\epsilon,\infty)}(t) \right) dt\right) \\
			&= 	\sigma C t^{H-1/2}\Ze_t d\We_t \\
		\ea$$
	is a Martingale process under $\Qe$.	
	Note that under the measure $\Qe$, we have
		$$
			\Ze_t
			=	
			S_0\exp\left\{\sigma C \int_0^t s^{H-1/2} d\We_s - \frac{C^2\sigma^2}{4H} t^{2H}\right\}
		$$
	and similarly,
		\beq\label{geoGMQ}
			\Se_t
			=
			S_0\exp\left\{r t + \sigma C \int_0^t s^{H-1/2} d\We_s - \frac{C^2\sigma^2}{4H} t^{2H}\right\}.
		\eeq
	Finally,
		$$
			\mathbb{E_{\Qe}}[\Se_t] 
			= 	
			\mathbb{E_{\Qe}}\left[S_0\exp\left\{r t 
						+ \sigma C \int_0^t s^{H-1/2} d\We_s
						- \frac{C^2\sigma^2}{4H} t^{2H}\right\}\right] \\
			= 	
			S_0e^{rt},
		$$
	using It\^{o} Isometry and the moment generating function. Therefore $\Qe$ is in fact a risk-neutral measure. 
	
	\subsection{Option pricing}
	
	Let $F(T)$ be the payoff of an option on an asset with price $(\Se_t)_{t\in[0,T]}$ for some $\epsilon>\delta(H)T$ at time $T>0$. Note that we assume that the underlying stock price follows $(\Se_t)$, NOT the original stock price process $(S_t)$. 
	Define
		$$
			E_t=\mathbb{E_{\Qe}}(B_T^{-1}F|\mathcal{F}_t^H).
		$$
	Then by the Martingale Representation Theorem (see \cite{shreve_stochastic_2004}), there exists an adapted process $(\phi_t)_{t\in[0,T]}$ such that
		$$
			dE_t=\phi_td\Ze_t.
		$$
	For each $\epsilon>\delta(H)T$, we get a $\Delta$-hedging portfolio given by $(\phi_t,\psi_t)_{t\in[0,T]}$, where $\phi_t$ is the number of shares of the risky asset and $\psi_t=E_t-\phi_t\Ze_t$ is the number of shares of the bond at time $t$. It can be easily verified that the portfolio is self-financing and replicating under the modified stock price process $(\Se_t)$. Then by the standard no-arbitrage argument (see, for instance, \cite{shreve_stochastic_2004}), the value of the option is equal to the value of the portfolio at every time $t\in[0,T]$, given by
		\beq\label{optionPrice}
			\ba
						F_t
				&= 	\phi_t\Se_t + \psi_tB_t \\
				&=	B_t\mathbb{E_{\Qe}}(B_T^{-1}F|\mathcal{F}_t).
			\ea
		\eeq
	Furthermore, we can find the corresponding Black-Scholes partial differential equation:
	\bprop\label{blackscholesPDE}
		Consider an option with underlying stock price $(\Se_t)_{t\in[0,\infty)}$ as defined in \eqref{SteDEF} that has payoff $F$ at time $T>0$. The value of the option at time $t\in[0,T]$ is given by $c_t=f(\Se_t,t)$, where $f(x,t)$ is the solution to the partial differential equation
		$$
			rf(x,t) = rxf_x(x,t) + f_t(x,t) + \frac{1}{2}\sigma^2C^2t^{2H-1}x^2f_{xx}(x,t)
		$$
		with terminal condition $f(x,T)=F$.
	\eprop
		\bprof
			The underlying stock price process $(\Se_t)_{t\in[0,\infty)}$ satisfies, by \eqref{SteDEF} and \Cref{GMediffusion},
			$$
				d\Se_t = \alpha(t)\Se_tdt + \sigma Ct^{H-1/2}\Se_t dW_t,
			$$
			where $\alpha(t)=\mu + \sigma c_\Psi(2H-1)t^{2H-2}M_t\mathds{1}_{[\epsilon,\infty)}(t)$. 
			Then using It\^{o}'s formula, we have
			\beq\ba\label{bsPDE1}
						df(\Se_t,t)
				=&	f_x(\Se_t,t)d\Se_t + f_t(\Se_t,t) + \frac{1}{2}f_{xx}(\Se_t,t)(d\Se_t)^2 \\
				=&	f_x(\Se_t,t)\left(\alpha(t)\Se_tdt + \sigma Ct^{H-1/2}\Se_t dW_t\right) 
						+ f_t(\Se_t,t) \\
						&+ \frac{1}{2}\sigma^2C^2t^{2H-1}(\Se_t)^2f_{xx}(\Se_t,t)dt \\
				=&	\alpha(t)\Se_tf_x(\Se_t,t)dt + \sigma Ct^{H-1/2}\Se_t f_x(\Se_t,t) dW_t
						+ f_t(\Se_t,t) \\
						&+ \frac{1}{2}\sigma^2C^2t^{2H-1}(\Se_t)^2f_{xx}(\Se_t,t)dt.
			\ea\eeq
			Since the hedging portfolio $(\phi_t,\psi_t)$ is self-financing and replicates the value of the option at every time $t\in[0,T]$, we also have
			\beq\ba\label{bsPDE2}
						df(\Se_t,t)
				=&	\phi_td\Se_t + \psi_tdB_t \\
				=&	\phi_t\left(\alpha(t)\Se_tdt + \sigma Ct^{H-1/2}\Se_t dW_t\right) + \psi_trB_tdt \\
				=&	\phi_t\alpha(t)\Se_tdt + \phi_t\sigma Ct^{H-1/2}\Se_t dW_t + r\left(f(\Se_t,t)-\phi_t\Se_t\right)dt, \\
			\ea\eeq
			where $(B_t)_{t\in[0,\infty)}$ is the bond price process.
			Setting equations \eqref{bsPDE1} and \eqref{bsPDE2} equal gives
			$$\ba
				&		\left(\sigma Ct^{H-1/2}\Se_t f_x(\Se_t,t) - \phi_t\sigma Ct^{H-1/2}\Se_t\right)dW_t \\
				=& 	\left(\phi_t\alpha(t)\Se_t+\psi_trB_t-\alpha(t)\Se_tf_x(\Se_t,t)-f_t(\Se_t,t)
						-\frac{1}{2}\sigma^2C^2t^{2H-1}(\Se_t)^2f_{xx}(\Se_t,t)\right)dt
			\ea$$
			Since the left-hand side of this equation is a martingale process and the right-hand side is not,  they must both be equal to zero almost surely. Therefore, the number of shares of the underlying stock in the replicating portfolio $(\phi_t)_{t\in[0,T]}$ satisfies
			$$
				\phi_t = f_x(\Se_t,t)
			$$
			and finally,
			$$
				rf(x,t) = rxf_x(x,t) + f_t(x,t) + \frac{1}{2}\sigma^2C^2t^{2H-1}x^2f_{xx}(x,t)
			$$
			as required.
		\eprof
		
	\subsection{Computation of a call option price}
		
	The payoff $F$ of a call option on a risky asset with price $(\Se_t)_{t\in[0,T]}$ that has strike price $K$ and expiration $T$ is given by
		$$
			F = (\Se_T-K)^+.
		$$
	Suppose also that we have a risk-free interest rate $r$.
	Therefore by \eqref{optionPrice} and \eqref{geoGMQ}, we have
		$$\ba
					F_t
			&=	B_t\mathbb{E_{\Qe}}(B_T^{-1}F|\mathcal{F}_t) \\
			&=	B_t\mathbb{E_{\Qe}}(B_T^{-1}(\Se_T-K)^+|\mathcal{F}_t) \\
			&=	B_t\mathbb{E_{\Qe}}\left(B_T^{-1}\left.\left(\Se_t\frac{\Se_T}{\Se_t}-K\right)^+ \right|\mathcal{F}_t\right) \\
			&=	e^{-r(T-t)}\mathbb{E_{\Qe}}\left(\left.\left(\Se_te^{r(T-t)
					+ \sigma C \int_t^T s^{H-1/2} d\We_s
					- \frac{1}{2}\sigma^2C^2\left(\frac{T^{2H}-t^{2H}}{2H}\right)}-K\right)^+\right|\mathcal{F}_t\right).
		\ea$$
	Since $\Se_t$ is measurable with respect to $\mathcal{F}_t$, fix $x=\Se_t$. Then since $\int_t^T s^{H-1/2}d\We_s$ is independent of $\mathcal{F}_t$, we have
		$$\ba
					F_t
			=&	e^{-r(T-t)}\mathbb{E_{\Qe}}\left(\left.\left(xe^{r(T-t)
					+ \sigma C \int_t^T s^{H-1/2} d\We_s
					- \frac{1}{2}\sigma^2C^2\left(\frac{T^{2H}-t^{2H}}{2H}\right)}-K\right)^+\right|x=\Se_t\right)
		\ea$$
	Since $\int_t^T s^{H-1/2} d\We_s$ is a centered Gaussian random variable with variance $\frac{T^{2H}-t^{2H}}{2H}$, we have
		$$
						F_t
				=	e^{-r(T-t)}\frac{1}{\sqrt{2\pi}}\int_{-\infty}^{\infty}
						\left(\Se_te^{r(T-t)
						+ \sigma C \sqrt{\frac{T^{2H}-t^{2H}}{2H}} z
						- \frac{1}{2}\sigma^2 C^2 \left(\frac{T^{2H}-t^{2H}}{2H}\right)}-K\right)^+
						e^{-\frac{1}{2}z^2}dz,
		$$
	where $Z$ is a standard normal random variable.
	We have
		$$
			\Se_te^{r(T-t) + \sigma C \sqrt{\frac{T^{2H}-t^{2H}}{2H}}z - \frac{1}{2}\sigma^2C^2\left(\frac{T^{2H}-t^{2H}}{2H}\right)}-K \geq 0
		$$
	when
		$$
			z \geq d_1 := \frac	{\ln\left(\frac{K}{\Se_t}\right)-r(T-t)+\frac{1}{2}\sigma^2C^2\left(\frac{T^{2H}-t^{2H}}{2H}\right)}
													{\sigma C\sqrt{\frac{T^{2H}-t^{2H}}{2H}}},
		$$
	and therefore
		$$
			\ba
						F_t
				&=	e^{-r(T-t)}\frac{1}{\sqrt{2\pi}}\int_{d_1}^{\infty}
						\left(\Se_te^{r(T-t)
						+ \sigma C \sqrt{\frac{T^{2H}-t^{2H}}{2H}} z
						- \frac{1}{2}\sigma^2 C^2\left(\frac{T^{2H}-t^{2H}}{2H}\right)}-K\right)
						e^{-\frac{1}{2}z^2}dz \\
				&=	\Se_t\frac{1}{\sqrt{2\pi}}\int_{d_1}^{\infty}e^{-\frac{1}{2}\left(z-\sigma C \sqrt{\frac{T^{2H}-t^{2H}}{2H}}\right)^2}dz
						- Ke^{-r(T-t)}\frac{1}{\sqrt{2\pi}}\int_{d_1}^{\infty}e^{-\frac{1}{2}z^2}dz \\
				&=	\Se_t\Phi\left(\sigma C \sqrt{\frac{T^{2H}-t^{2H}}{2H}}-d_1\right)
						- Ke^{-r(T-t)}\Phi(-d_1).
			\ea
		$$
	We observe that when $H=1/2$, this formula is consistent with the original Black-Scholes call option price.
\section{Parameter estimation techniques}\label{Parameter estimation techniques}

	In both the original Black-Scholes model, its analogue with fractional Brownian motion, and now the model with the Dobri\'{c}-Ojeda process as the driving noise for the stock price process, we assume that the stock price parameters $\mu$, $\sigma$, and $H$ (drift, volatility, and Hurst index, respectively) are constant for $t\in[0,T]$. In this section we discuss two methods for estimating these parameters based on historical stock price data.

\subsection{Ratio method with Ergodic Theory}\label{sec:paramsfbm}

	First, we examine a parameter estimation technique developed in \cite{scansaroli_stochastic_2012}. In order to employ this technique, we will assume that the Hurst index $H$ of the stock price following a geometric Dobri\'{c}-Ojeda process is the same parameter $H$ of the corresponding geometric fractional Brownian motion process, i.e. we assume that $H_{Z_H}=H_{\Ve_H}$. We justify this assumption by noting that the processes $(Z_H(t))$ and $(V_H(t))$ behave similarly, with less than 12\% relative error, as discussed in \Cref{IntroductionChapter}. Under this assumption, we can employ the stationary and ergodic properties of the increments of fractional Brownian motion in a ratio method for estimating $H$.
	
	Define the shift transformation $\tau$ on a stochastic process $\{Y(t)\}_{t\geq0}$ by $(Y\circ\tau)(t)=Y(t+\Delta t)-Y(\Delta t)$ for some small fixed $\Delta t$. Next define the sequence of random variables $\{X_m\}_{m\in\mathds{Z}^+}$ by $X_m = Z_H\circ\tau^m$, where $Z_H$ is a fractional Brownian motion process. The process $Z_H$ is invariant in law with respect to a shift $\tau$ in time, since $(Z_H\circ\tau^m)(t)=Z_H(t+m\Delta t)-Z_H(m\Delta t)$ and fractional Brownian motion has stationary increments. Thus the sequence $\{X_m\}$ is ergodic.
	
	Therefore, by the ergodic theorem (see \cite[p~337,Thm~2.1]{durrett_probability:_2004}), the sum of increments of fractional Brownian motion converge to their mean, 0, and the sum of squared increments of fractional Brownian motion converge to their second moment. We will use this fact to estimate the parameters $\mu$, $\sigma$, and $H$.
	
	Suppose that $s_i$ is the observed price of the underlying stock at time $t_i=\frac{iT}{n}$, for $i=0,\ldots,n$. Note that the time between each observation, $\Delta t=\frac{T}{n}$, is fixed. For example, $s_i$ may be daily closing prices. Without loss of generality, assume that the stock does not pay dividends during the interval $[0,T]$. Otherwise use the adjusted stock price. Define the log returns $y_i=\ln\frac{s_i}{s_{i-1}}$ for $i=1,\ldots,n$. Then under the assumption that the stock price follows a geometric fractional Brownian motion process as in \eqref{geofBM},
set
	$$
		y_i = \mu\Delta t + \sigma(Z_H(t_i)-Z_H(t_{i-1})) - \frac{1}{2}\sigma^2(t_i^{2H}-t_{i-1}^{2H}).
	$$
Then we have
	\beq\ba\label{eq:riemannsum1}
				\frac{1}{n}\sum_{i=1}^n \frac{1}{2}\sigma^2(t_i^{2H}-t_{i-1}^{2H})
		=&	\frac{\sigma^2}{2n}\left(\frac{T}{n}\right)^{2H}\sum_{i=1}^n (i^{2H}-(i-1)^{2H}) \\
		=&	\frac{H\sigma^2}{n}\left(\frac{T}{n}\right)^{2H}\sum_{i=1}^n (x_i^*)^{2H-1},
	\ea\eeq
for some $x_i^*\in[i-1,i]$, $i=1,\ldots,n$. Since $\frac{x_i^*}{n}\in[\frac{i-1}{n},\frac{i}{n}]$, we can rearrange to achieve a Riemann sum:
	\beq\label{eq:riemannsum}
		\frac{H\sigma^2T^{2H}}{n}\sum_{i=1}^n \left(\frac{x_i^*}{n}\right)^{2H-1}\frac{1}{n}
	\eeq
The sum converges to $\int_0^1 x^{2H-1}\,dx=\frac{1}{2H}$ as $n\rightarrow\infty$ and thus equation \eqref{eq:riemannsum} converges to 0 as $n\rightarrow\infty$.
By using the ergodic property of $(Z_H(t))$, we have
	\beq\label{eq:deltaZhto0}
		\frac{1}{n}\sum_{i=1}^n(Z_H(t_i)-Z_H(t_{i-1})) \rightarrow \mathbb{E}(Z_H(t_1)-Z_H(t_0)) = 0 \text{ a.s.}
	\eeq
and so
	$$
		\frac{1}{n}\sum_{i=1}^n y_i = \mu\Delta t + \frac{\sigma}{n}\sum_{i=1}^n(Z_H(t_i)-Z_H(t_{i-1})) - \frac{1}{n}\sum_{i=1}^n\frac{1}{2}\sigma^2(t_i^{2H}-t_{i-1}^{2H}) 
		\rightarrow \mu\Delta t.
	$$
Therefore we will estimate the drift $\mu$ for $n$ sufficiently large by
	$$
		\mu \approx \hat{\mu} = \frac{1}{\Delta t}\frac{1}{n}\sum_{i=1}^n y_i.
	$$
Since it remains to estimate both the volatility $\sigma$ and the Hurst index $H$, we will use a ratio of second moments to estimate $H$ first, as in \cite{scansaroli_stochastic_2012}. Using the previously computed estimator $\hat{\mu}$, let
	$$\ba
		SS_1 	:=& \frac{1}{n}\sum_{i=1}^n (y_i-\hat{\mu}\Delta t)^2 \\
					=&	\frac{1}{n}\sum_{i=1}^n \left(\sigma(Z_H(t_i)-Z_H(t_{i-1})) - \frac{1}{2}\sigma^2(t_i^{2H}-t_{i-1}^{2H})\right)^2 \\
					=&	\frac{\sigma^2}{n}\sum_{i=1}^n(Z_H(t_i)-Z_H(t_{i-1}))^2 
							- \frac{\sigma^3}{n}\sum_{i=1}^n(Z_H(t_i)-Z_H(t_{i-1})(t_i^{2H}-t_{i-1}^{2H}) \\
							&+ \frac{\sigma^4}{4n}\sum_{i=1}^n(t_i^{2H}-t_{i-1}^{2H})^2 \\
					\longrightarrow& \, \sigma^2(\Delta t)^{2H}.
	\ea$$
The first term converges to $\sigma^2(\Delta t)^{2H}$ as $n\rightarrow\infty$ since by the ergodic theorem, $1/n$ times the sum of the squared increments of $Z_H$ converges to the increments' second moment. The third term converges to 0 since as shown in \eqref{eq:riemannsum1} and \eqref{eq:riemannsum}, the function $f(x)=x^{2H}$ has finite variation so it must have quadratic variation 0. The second term converges to 0 since by the Cauchy-Schwarz inequality, it can be written as the square root of the product of the first and third terms. 
To achieve a ratio for our estimator, we define $SS_2$ by sampling half as many points as in $SS_1$: Let
	$$
		SS_2 := \frac{1}{\lfloor n/2\rfloor}\sum_{i=1}^{\lfloor n/2 \rfloor} \left(\ln\frac{s_{2i}}{s_{2i-1}}-\hat{\mu}(2\Delta t)\right)^2
					\rightarrow \sigma^2(2\Delta t)^{2H},
	$$
by the same computation as in $SS_1$, above.
Then since all convergence is almost sure, we can take the quotient:
	$$
		\frac{SS_1}{SS_2} \rightarrow \left(\frac{1}{4}\right)^H
	$$
and thus estimate the Hurst index $H$ by
	$$
		H \approx \hat{H} = \log_4\left(\frac{SS_1}{SS_2}\right).
	$$
Finally, we can use $\hat{\mu}$ and $\hat{H}$ to estimate the volatility $\sigma$:
	$$
		\sigma^2 \approx \hat{\sigma}^2 = \frac{1}{(\Delta t)^{2\hat{H}}}\frac{1}{n}\sum_{i=1}^n (y_i-\hat{\mu}\Delta t)^2.
	$$
		
\subsection{Parameter estimation using quadratic variation}\label{sec:paramsGM}
	
	Next we relax the assumption that the parameters of the Dobri\'{c}-Ojeda model are necessarily equal to the parameters of the fractional Brownian motion model, i.e. that $H_{Z_H}=H_{\Ve_H}$. We aim to estimate $H$ and $\sigma$ using properties of the modified Dobri\'{c}-Ojeda process. (The drift $\mu$ plays no role in pricing an option so we omit its estimation.) Unlike fractional Brownian motion, the modified Dobri\'{c}-Ojeda process does not have ergodic increments so we cannot use the technique described in \Cref{sec:paramsfbm}. Therefore, we propose the use of quadratic variation to estimate parameters in this model.
		
	\subsubsection{Almost-sure convergence of the quadratic variation}

		First, recall the definition of quadratic variation:
		\bdfn
			Let $f(t)$ be a function defined on the interval $[t_0,T]$. The quadratic variation of $f$ from time $t_0$ to time $T$, $\tensor*[_{t_0}]{[f,f]}{_T}$, is defined as
				$$
					\tensor*[_{t_0}]{[f,f]}{_T} = \lim_{||\Pi_n||\rightarrow 0}\sum_{j=1}^n (f(t_j)-f(t_{j-1}))^2
				$$
			where $\Pi_n = \{t_0,t_1,\ldots ,t_n\}$, $t_0<t_1<\ldots<t_n = T$ and $||\Pi_n||=\max_{j=1,\ldots,n}(t_j-t_{j-1})$.
		\edfn
		As shown in 
		\Cref{GMQV,GMeQV}, the quadratic variation of both the original Dobri\'{c}-Ojeda process $(V_H(t))$ and the modified Dobri\'{c}-Ojeda process $(\Ve_H(t))$ are given by
			$$I := \tensor*[_{t_0}]{[\Ve_H,\Ve_H]}{_T} = \frac{C^2}{2H} (T^{2H}-t_0^{2H}).$$
		We use the following theorem to construct a parameter technique that uses the quadratic variation of $(\Ve_H(t))$. We will prove convergence in $L^2$, where the $L^2$ norm, $||\cdot||_2$ is given by
			$$
				||X||_2 = \sqrt{\mathbb{E}[X^2]},
			$$
		and also almost sure convergence, which will allow us to use another ratio method to estimate the Hurst index, $H$. We require a sampling rate strictly greater than $n$ in order to ensure almost sure convergence.
		\bthm\label{GMQVthm}
			Let $t_i=\tfrac{iT}{\lfloor n^{1+\delta}\rfloor}$,
			$i=i_0,...,\lfloor n^{1+\delta} \rfloor$ ,
			$i_0=\tfrac{t_0\lfloor n^{1+\delta}\rfloor}{T}$,
			be a sequence of partitions of $[t_0,T]$ for some $\delta>0$ and $V_t=V_H(t)$ as defined in \eqref{GMdef}. 
			Then
				$$\lim_{n\rightarrow\infty}\left|\left|\sum\limits_{i=i_0}^n{(\Delta V_{t_i})^2} 
					- I\right|\right|_2 = 0$$
			and
				$$\lim_{n\rightarrow\infty}\sum\limits_{i=i_0}^{\lfloor n^{1+\delta} \rfloor}{(\Delta V_{t_i})^2} = I 
					\qquad \text{a.s.}$$
			where $\Delta V_{t_i}=V_{t_i}-V_{t_{i-1}}$.
		\ethm
	To achieve almost sure convergence, we need to sample at a rate strictly greater than $n$, or $\lfloor n^{1+\delta}\rfloor$, for $\delta>0$. In practice, this only impacts the precision of our estimator. Please see \Cref{sec:appendix} for proof of \Cref{GMQVthm}.
		\bcor
			The sample quadratic variation of the modified Dobri\'{c}-Ojeda process $(\Ve_t)$ converges in $L^2$ and almost surely to $I=\frac{C^2}{2H} (T^{2H}-t_0^{2H})$.
		\ecor
			\bprof
				As the only modification to the original Dobri\'{c}-Ojeda process is in the drift term and the drift term does not impact quadratic variation, the quadratic variation remains unchanged.
			\eprof
		Now we define the log of the stock price process, $X_t=\ln(\Se_t)$. Then we also have convergence of the quadratic variation of $X_t$:
		\bcor\label{StQVthm}
			The sample quadratic variation of the log stock price process $(X_t)$ converges in $L^2$ and almost surely to $\frac{C^2\sigma^2}{2H} (T^{2H}-t_0^{2H})$.
		\ecor
			\bprof
				As in \eqref{geoGM}, we can write $X_t$ as
				$$
					X_t = \ln(\Se_t) = \mu t + \sigma\Ve_t - \frac{C^2\sigma^2}{4H}t^{2H}
				$$
				and again, since the only difference between $X_t$ and $\Ve_t$ is in the drift, the quadratic variation is simply $\sigma^2 I$.
			\eprof

	\subsubsection{Ratio method with quadratic variation}

	As in \Cref{sec:paramsfbm}, suppose we have $m=\lfloor n^{1+\delta}\rfloor$ equally time-spaced observations of the stock price process $(\Se_t)$, called $s_i$, observed at time $t_i=\frac{iT}{m}$, $i=0,\ldots,m$. Let $\Delta t = \tfrac{T}{m}$. Again, assume that the stock price does not pay dividends during this interval and define the log returns $y_i=\ln\frac{s_i}{s_{i-1}}$ for $i=1,\ldots,m$. We assume the stock price process follows a geometric Dobri\'{c}-Ojeda process, as detailed in \Cref{sec:DOandoptionpricing}, where we have
		$$
			\Se_t = S_0\exp\left\{r t + \sigma C \int_0^t s^{H-1/2} d\We_s - \frac{\sigma^2 C^2}{2(2H)} t^{2H}\right\}.
		$$
	Assume
		$$
			y_i = \mu\Delta t + \sigma (\Ve_H(t_i)-\Ve_H(t_{i-1})) - \frac{1}{2}\sigma^2\frac{C^2}{2H}(t_i^{2H}-t_{i-1}^{2H}).
		$$
	By \Cref{StQVthm}, we have
		$$
			\sum_{i=1}^m y_i^2 
			\rightarrow \sigma^2\frac{C^2}{2H}T^{2H}
		$$
	and similarly, the sample quadratic variation of half of the sample path converges:
		$$
			\sum_{i=1}^{\lfloor m/2 \rfloor} y_i^2 
			\rightarrow \sigma^2\frac{C^2}{2H}\left(\frac{T}{2}\right)^{2H}.
		$$
	Therefore, since this convergence is almost sure, we can use a ratio of quadratic variations method to estimate the parameter $H$:
		$$
			\frac{\sum_{i=1}^{\lfloor m/2 \rfloor} y_i^2 }
				{\sum_{i=1}^m y_i^2}
			\rightarrow
			\frac{\sigma^2\frac{C^2}{2H}\left(\frac{T}{2}\right)^{2H}}
				{\sigma^2\frac{C^2}{2H}T^{2H}}
			= \left(\frac{1}{4}\right)^H.
		$$
	Therefore for $m$ sufficiently large, we will estimate the Hurst index $H$ by
		$$
			H 
			\approx \hat{H}
			= \log_4\left(\frac{\sum_{i=1}^{\lfloor m/2 \rfloor} y_i^2 }
				{\sum_{i=1}^m y_i^2}\right).
		$$
	Finally, we can use the estimator $\hat{H}$ to obtain an estimate for the volatility $\sigma$:
		$$
			\sigma^2
			\approx \hat{\sigma}^2
			= \frac{2}{C\hat{H}T^{2\hat{H}}}\sum_{i=1}^m y_i^2.
		$$
\section{Simulation and case study}

We conclude the development of this model with a brief mention of simulation and finally computation of the value of a European call option using historical stock price data.

\subsection{Simulation}
Using the It\^{o} diffusion representation of the Dobri\'{c}-Ojeda process given in \Cref{GMdiffusion}, we can use a sequence of i.i.d. standard normal random variables in order to simulate a discretized Dobri\'{c}-Ojeda sample path, assuming that $V_H(0)=0$. 
More specifically, if $\{X_i\}_{i=1,\ldots,n}$ is a sequence of i.i.d. standard normal random variables, then we simulate increments $\Delta M_H(t_i)$ of the martingale process by
$$
	\Delta M_H(t_i) =\sqrt{c_M(2-2H)}t_i^{1/2-H}\sqrt{\Delta t} X_i.
$$
We sum the increments $\Delta M_H(t_i)$ and multiply by the deterministic function $\Psi_H(t)$ to simulate a sample path of $V_H(t)$.

To describe implementation of the model, we price a historical European call option and compare this price with the actual trading price along with prices computed using the original Black-Scholes model and the model using fractional Brownian motion as its driving process, as developed by Hu and Oksendal \cite{hu_fractional_2003} and Sottinen \cite{sottinen_fractional_2001}.

\subsection{Case study: AAL}\label{sec:AAL}
\begin{figure}[htbp]
  \centering
	\includegraphics[width=\textwidth,keepaspectratio=true]{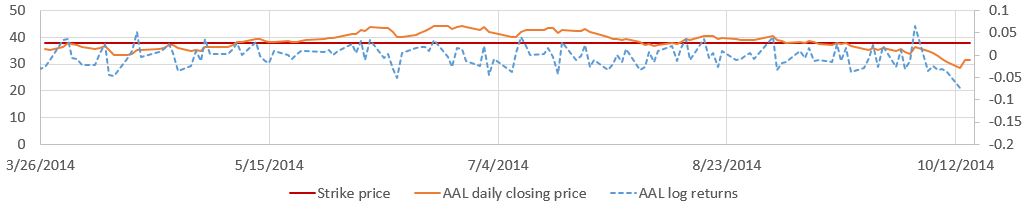}
  \caption{Graph of AAL daily closing prices.}
  \label{fig:AALlogreturns}
\end{figure}

\begin{figure}[htbp]
  \centering
	\includegraphics[width=\textwidth,keepaspectratio=true]{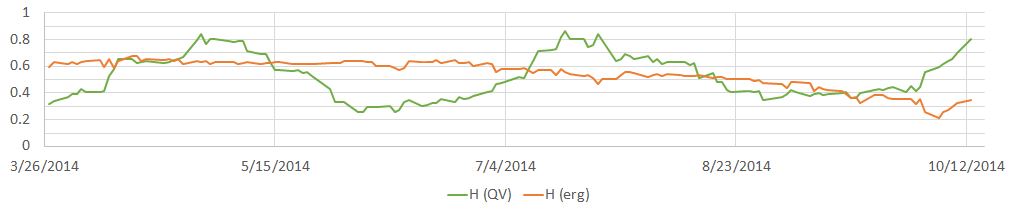}
  \caption{Graph of rolling $H$ estimates.}
  \label{fig:AALh2}
\end{figure}

\begin{figure}[htbp]
  \centering
	\includegraphics[width=\textwidth,keepaspectratio=true]{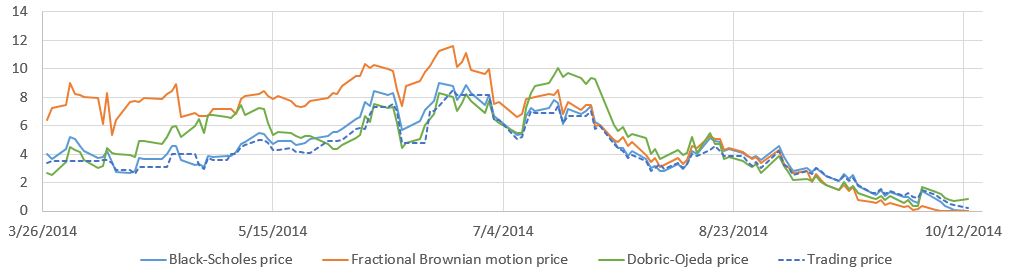}
  \caption{Graph of computed option prices.}
  \label{fig:AALoption4}
\end{figure}

	We consider a call option on American Airlines stock (AAL) with strike price $K=38$ and expiration November 22, 2014. For each day beginning March 27, 2014 and ending October 15, 2014, we estimate $H$ and $\sigma$ using the previous 62 consecutive daily AAL closing prices. \Cref{fig:AALlogreturns} shows the daily closing price for the stock over this time period. 
	For each day, we compute 3 estimations for the parameters: 1. assuming the stock price follows a geometric Brownian motion process and using standard Black-Scholes techniques; 2. assuming the stock price follows a geometric fractional Brownian motion process and using a ratio of second moments as detailed in \Cref{sec:paramsfbm}; 3. assuming the stock price follows a geometric Dobri\'{c}-Ojeda process and using a ratio of quadratic variations, as detailed in \Cref{sec:paramsGM}. The latter two rolling estimates for $H$ are shown in \Cref{fig:AALh2}.
	One immediate observation is that the estimate of $H$ using quadratic variation is extremely sensitive to large changes in the log return of the underlying stock. We also notice that the estimates for $H$ are in both cases often significantly lower than 0.6, our market-wide expected $H$ estimate discussed in \Cref{IntroductionChapter}. These observations lead us to believe that $H$ varies both over time and over stock selection. Next we compute the option price using the three competing models and their respective parameter estimation techniques and compare these prices to the actual trading price of the stock at market close each day. The results are shown in \Cref{fig:AALoption4}. 
	
\subsection{Case study: BAC}
\begin{figure}[htbp]
  \centering
	\includegraphics[width=\textwidth,keepaspectratio=true]{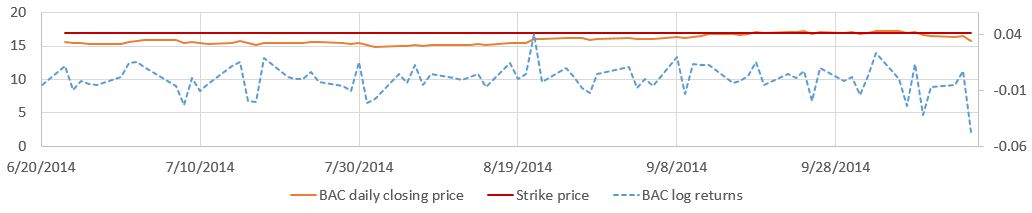}
  \caption{Graph of BAC daily closing prices.}
  \label{fig:BAClogreturns}
\end{figure}

\begin{figure}[htbp]
  \centering
	\includegraphics[width=\textwidth,keepaspectratio=true]{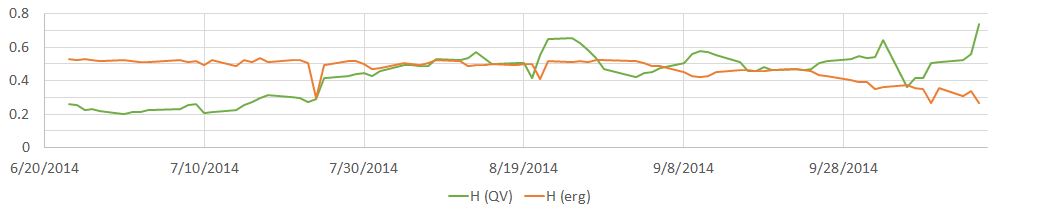}
  \caption{Graph of rolling $H$ estimates.}
  \label{fig:BACh2}
\end{figure}

\begin{figure}[htbp]
  \centering
	\includegraphics[width=\textwidth,keepaspectratio=true]{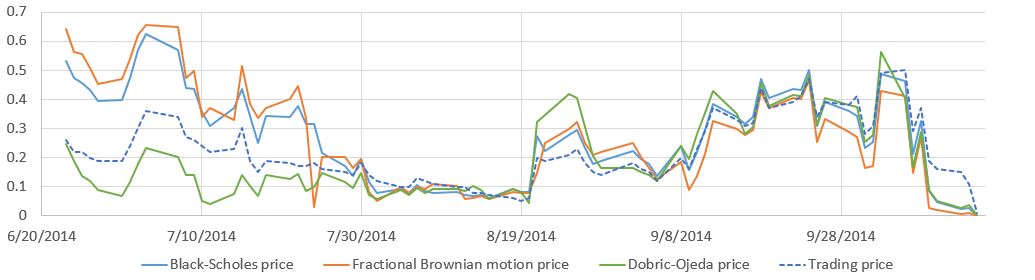}
  \caption{Graph of computed option prices.}
  \label{fig:BACoption4}
\end{figure}

	We consider a call option on Bank of America stock (BAC) with strike price $K=17$ and expiration October 18, 2014. For each day beginning June 23, 2014 and ending October 15, 2014, we estimate $H$ and $\sigma$ using the previous 62 consecutive daily BAC closing prices. \Cref{fig:BAClogreturns} shows the daily closing price for the stock over this time period. 
	All estimators are computed as in \Cref{sec:AAL}. The two rolling estimates of $H$ are shown in \Cref{fig:BACh2}.
	We notice that when the $H$ estimates using fractional Brownian motion and the Dobri\'{c}-Ojeda process are similar, the two models' computed option prices are also similar, as expected. We also notice that when our ratio of quadratic variations method yields a lower value of $H$, the Dobri\'{c}-Ojeda option price is more accurate with respect to the actual trading price of the option than both the Black-Scholes price and the fractional Brownian motion price using a higher $H$ parameter.
	Next we compute the option price using the three competing models and their respective parameter estimation techniques and compare these prices to the actual trading price of the stock at market close each day. The results are shown in \Cref{fig:BACoption4}. 
	
	In general, when the quadratic variation method yields a higher value for $H$, the Dobri\'{c}-Ojeda model correspondingly overestimates the option price. However, when the $H$ estimate using $V_H(t)$ is lower than expected, this model outperforms the others in approximating the actual trading price of the option. We also notice (less surprisingly) that the Black-Scholes price is fairly similar to the option's trading price. A more accurate method of testing the various models would be in building competing virtual historical portfolios and considering their performance.
	
	\appendix

\section{Proof of \Cref{GMQVthm}}

We will utilize the following lemma in the proof of \Cref{GMQVthm}.
		\blem\label{I*lemma}
			For
			\begin{equation}\label{I*def}
				I^* = C^2
			 \sum\limits_{i=i_0}^{\lfloor n^{1+\delta}\rfloor}t_i^{2H-1}\Delta t,
			\end{equation}
			where $\Delta t = t_i-t_{i-1}$,
			we have
				$$\lim_{n\rightarrow\infty}{I^*} = I \qquad \text{a.s.}$$
		\elem

		\bprof
			We have
			$$
				I = \int_{t_0}^t C^2 s^{2H-1}ds = \lim_{n\rightarrow\infty}\sum_{i=i_0}^{\lfloor n^{1+\delta}\rfloor} C^2 t_i^{2H-1}\Delta t,
			$$
			by the definition of a definite Riemann integral.
		\eprof


	\bprof[Proof of \Cref{GMQVthm}.]
		Let $m=\lfloor n^{1+\delta} \rfloor$. By the triangle inequality, 
			$$\left|\left|\sum\limits_{i=i_0}^m (\Delta V_{t_i})^2 -I\right|\right|_2 
				\leq 
				\left|\left|\sum\limits_{i=i_0}^m (\Delta V_{t_i})^2 -I^*\right|\right|_2 + ||I^*-I||_2$$ 
		and so by \Cref{I*lemma}, it suffices to show that 
			$\left|\left|\sum\limits_{i=i_0}^m (\Delta V_{t_i})^2 -I^*\right|\right|_2 \longrightarrow 0$.
		We have, by \eqref{I*def},
		$${I^*}^2 
			= C^4
				\sum\limits_{i=i_0}^m
				\sum\limits_{j=i_0}^m t_i^{2H-1}t_j^{2H-1} (\Delta t)^2. $$
		We will need the approximations for $M_t$ given in \Cref{DeltaMapprox}.
		Similarly, we can approximate $\Delta\Psi_{t_i}$ and $(\Delta\Psi_{t_i})^2$:
			$$
				\Delta\Psi_{t_i}
				= c_\Psi(t_i^{2H-1}-t_{i-1}^{2H-1})
				\approx c_\Psi(2H-1)t_i^{2H-2}\Delta t
			$$
		and
			$$
				(\Delta\Psi_{t_i})^2
				\approx c_\Psi^2(2H-1)^2t_i^{4H-4}(\Delta t)^2.
			$$
		Then we have
		\beq\label{allthreeterms}
									\left|\left|\sum\limits_{i=i_0}^m (\Delta V_{t_i})^2 -I^*\right|\right|_2^2
				=					\mathbb{E}\left[\sum\limits_{i=i_0}^m \sum\limits_{j=i_0}^m (\Delta V_{t_i})^2 (\Delta V_{t_j})^2 \right] 
										- 2I^*	\mathbb{E}\left[\sum\limits_{i=i_0}^m(\Delta V_{t_i})^2\right] 
										+ {I^*}^2.
		\eeq
		Note that we can write $(\Delta V_{t_i})^2$ as
		$$
			(\Psi_{t_i}\Delta M_{t_i}+\Delta\Psi_{t_i}M_{t_{i-1}})^2 = (\Delta \Psi_{t_i})^2 M_{t_{i-1}}^2 + 2\Delta\Psi_{t_i}M_{t_{i-1}}\Psi_{t_i}\Delta M_{t_i} + \Psi_{t_i}^2(\Delta M_{t_i})^2,
		$$
		so the last two terms of \eqref{allthreeterms} give
		$$\ba
			&		- 2I^*\mathbb{E}\left[\sum\limits_{i=i_0}^m(\Delta V_{t_i})^2 \right] + {I^*}^2 \\
			=&	- 2C^2\sum\limits_{j=i_0}^mt_j^{2H-1}\Delta t
						\sum\limits_{i=i_0}^m\left((\Delta \Psi_{t_i})^2 \mathbb{E}[M_{t_{i-1}}^2] 
																				+ 2\Delta\Psi_{t_i}\Psi_{t_i}\mathbb{E}[M_{t_{i-1}}\Delta M_{t_i}] \right. 
												+ \left. \Psi_{t_i}^2\mathbb{E}[(\Delta M_{t_i})^2]\right) \\
					&+ C^4\sum\limits_{i=i_0}^m\sum\limits_{j=i_0}^m t_i^{2H-1}t_j^{2H-1} \Delta t^2. \\
		\ea$$
		Using the above estimations, this is approximately equal to
		\beq\ba\label{lasttwoterms}
				-& 2C^2\sum\limits_{j=i_0}^mt_j^{2H-1}\Delta t
						\sum\limits_{i=i_0}^m\left(c_Mc_\Psi^2(2H-1)^2t_i^{4H-4}(\Delta t)^2 t_i^{2-2H} \right. \\
						&+ \left. c_M(2-2H)c_\Psi^2t_i^{4H-2}t_i^{1-2H}\Delta t\right)
						+ C^4\sum\limits_{i=i_0}^m\sum\limits_{j=i_0}^m t_i^{2H-1}t_j^{2H-1} (\Delta t)^2 \\
			=&	\sum\limits_{j=i_0}^m \sum\limits_{i=i_0}^m
						\left(- 2C^2c_Mc_\Psi^2(2H-1)^2 t_i^{2H-2}t_j^{2H-1} (\Delta t)^3
									- C^4t_i^{2H-1}t_j^{2H-1} (\Delta t)^2\right).
		\ea\eeq
		We will see that the first term of \eqref{lasttwoterms} converges and the second term, 
		$$ -\sum\sum C^4t_i^{2H-1}t_j^{2H-1} (\Delta t)^2, $$
		is canceled by another term.
		The first term of \eqref{allthreeterms} is slightly less enjoyable to compute:
		$$\ba
			&		\sum\limits_{i=i_0}^m \sum\limits_{j=i_0}^m \mathbb{E}[(\Delta V_{t_i})^2 (\Delta V_{t_j})^2] \\
			=&	\sum\limits_{i=i_0}^m \sum\limits_{j=i_0}^m \mathbb{E}[
					((\Delta \Psi_{t_i})^2 M_{t_{i-1}}^2 + 2\Delta\Psi_{t_i}M_{t_{i-1}}\Psi_{t_i}\Delta M_{t_i} + \Psi_{t_i}^2(\Delta M_{t_i})^2) \\
					&\cdot
					((\Delta \Psi_{t_j})^2 M_{t_{j-1}}^2 + 2\Delta\Psi_{t_j}M_{t_{j-1}}\Psi_{t_j}\Delta M_{t_j} + \Psi_{t_j}^2(\Delta M_{t_j})^2)
					] \\
			=&	\sum\limits_{i=i_0}^m \sum\limits_{j=i_0}^m \left[
					(\Delta \Psi_{t_i})^2 (\Delta \Psi_{t_j})^2 \mathbb{E}[M_{t_{i-1}}^2M_{t_{j-1}}^2]
					+ 2(\Delta \Psi_{t_i})^2 \Delta\Psi_{t_j}\Psi_{t_j}\mathbb{E}[M_{t_{i-1}}^2M_{t_{j-1}}\Delta M_{t_j}]\right. \\
					&+ (\Delta \Psi_{t_i})^2 \Psi_{t_j}^2\mathbb{E}[M_{t_{i-1}}^2(\Delta M_{t_j})^2]
					+ 2\Delta\Psi_{t_i}\Psi_{t_i}(\Delta \Psi_{t_j})^2 \mathbb{E}[M_{t_{i-1}}\Delta M_{t_i}M_{t_{j-1}}^2] \\
					&+ 4\Delta\Psi_{t_i}\Psi_{t_i}\Delta\Psi_{t_j}\Psi_{t_j}\mathbb{E}[M_{t_{i-1}}\Delta M_{t_i}M_{t_{j-1}}\Delta M_{t_j}] \\
					&+ 2\Delta\Psi_{t_i}\Psi_{t_i}\Psi_{t_j}^2\mathbb{E}[M_{t_{i-1}}\Delta M_{t_i}(\Delta M_{t_j})^2]\\
					&+ \Psi_{t_i}^2(\Delta \Psi_{t_j})^2 \mathbb{E}[(\Delta M_{t_i})^2M_{t_{j-1}}^2]
					+ 2\Psi_{t_i}^2\Delta\Psi_{t_j}\Psi_{t_j}\mathbb{E}[(\Delta M_{t_i})^2M_{t_{j-1}}\Delta M_{t_j}] \\
					&+ \left. \Psi_{t_i}^2\Psi_{t_j}^2\mathbb{E}[(\Delta M_{t_i})^2(\Delta M_{t_j})^2] \right].\\
		\ea$$
		By symmetry, this is equal to
		$$\ba
			&		2\sum\limits_{i=i_0}^m \sum\limits_{i<j} \left[
					(\Delta \Psi_{t_i})^2 (\Delta \Psi_{t_j})^2 \mathbb{E}[M_{t_{i-1}}^2M_{t_{j-1}}^2]
					+ 2(\Delta \Psi_{t_i})^2 \Delta\Psi_{t_j}\Psi_{t_j}\mathbb{E}[M_{t_{i-1}}^2M_{t_{j-1}}\Delta M_{t_j}] \right. \\
					&+ (\Delta \Psi_{t_i})^2 \Psi_{t_j}^2\mathbb{E}[M_{t_{i-1}}^2(\Delta M_{t_j})^2]
					+ 2\Delta\Psi_{t_i}\Psi_{t_i}(\Delta \Psi_{t_j})^2 \mathbb{E}[M_{t_{i-1}}\Delta M_{t_i}M_{t_{j-1}}^2] \\
					&+ 4\Delta\Psi_{t_i}\Psi_{t_i}\Delta\Psi_{t_j}\Psi_{t_j}\mathbb{E}[M_{t_{i-1}}\Delta M_{t_i}M_{t_{j-1}}\Delta M_{t_j}] 
					+ 2\Delta\Psi_{t_i}\Psi_{t_i}\Psi_{t_j}^2\mathbb{E}[M_{t_{i-1}}\Delta M_{t_i}(\Delta M_{t_j})^2]\\
					&+ \Psi_{t_i}^2(\Delta \Psi_{t_j})^2 \mathbb{E}[(\Delta M_{t_i})^2M_{t_{j-1}}^2]
					+ 2\Psi_{t_i}^2\Delta\Psi_{t_j}\Psi_{t_j}\mathbb{E}[(\Delta M_{t_i})^2M_{t_{j-1}}\Delta M_{t_j}] \\
					&+ \left. \Psi_{t_i}^2\Psi_{t_j}^2\mathbb{E}[(\Delta M_{t_i})^2(\Delta M_{t_j})^2] \right]\\
					&+ \sum\limits_{i=i_0}^m \left[
					(\Delta \Psi_{t_i})^4 \mathbb{E}[M_{t_{i-1}}^4]
					+ 4(\Delta \Psi_{t_i})^3\Psi_{t_i}\mathbb{E}[M_{t_{i-1}}^3\Delta M_{t_i}] \right. \\
					&+ \left. 6(\Delta \Psi_{t_i})^2 \Psi_{t_i}^2\mathbb{E}[M_{t_{i-1}}^2(\Delta M_{t_i})^2] 
					+ 4\Delta\Psi_{t_i}\Psi_{t_i}^3\mathbb{E}[M_{t_{i-1}}(\Delta M_{t_i})^3] 
					+ \Psi_{t_i}^4\mathbb{E}[(\Delta M_{t_i})^4] \right].\\
		\ea$$
		We generalize the cross terms as follows:
		$$\ba
			(\Delta\Psi_{t_i})^\beta\Psi_{t_i}^{4-\beta}\mathbb{E}[M_{t_{i-1}}^\beta(\Delta M_{t_i})^{4-\beta}],
		\ea$$
		for $\beta=1,2,3,4$.
		The only nonzero cross terms correspond to $\beta=0,2,4$:
		$$\ba
			& (\Delta \Psi_{t_i})^4 \mathbb{E}[M_{t_{i-1}}^4]
			\approx	c_\psi^4(2H-1)^4t_i^{8H-8}(\Delta t)^4 c_M^2 t_i^{4-4H}
			= c_\Psi^4 c_M^2 (2H-1)^4 t_i^{4H-4}(\Delta t)^4, \\
			& 6(\Delta \Psi_{t_i})^2 \Psi_{t_i}^2\mathbb{E}[M_{t_{i-1}}^2(\Delta M_{t_i})^2]
			\approx 6c_\Psi^4c_M^2(2H-1)^2(2-2H)t_i^{4H-3}(\Delta t)^3 \text{, and}\\
			& \Psi_{t_i}^4\mathbb{E}[(\Delta M_{t_i})^4]
			\approx c_\Psi^4c_M^2(2-2H)^2 t_i^{4H-2}(\Delta t)^2.
		\ea$$
		To see that each of these terms converges, we compute in general, $\sum\limits_{i=i_0}^m t_i^{4H-K}(\Delta t_i)^K$ for $K\geq2$.
		Setting $t_i=\tfrac{iT}{m}$,
		we have
		$$\ba 
						 \sum\limits_{i=i_0}^m t_i^{4H-K}(\Delta t)^K 
					&= \left(\tfrac{T}{m}\right)^{4H} \sum\limits_{i=i_0}^m i^{4H-K} \\
					&= \left(\tfrac{T}{m}\right)^{4H} \left[i_0^{4H-K} + \sum\limits_{i=i_0+1}^m i^{4H-K}\right] \\
					&  \leq\left(\tfrac{T}{m}\right)^{4H} \left[i_0^{4H-K} + \int_{i_0}^m \! x^{4H-K} \, \mathrm{d}x.\right] \\
					&= T^{4H}
							\left[
								\frac{\left(\tfrac{t_0}{T}\right)^{4H-K}}{\lfloor n^{1+\delta}\rfloor^K}
								+ \tfrac{1}{4H-K+1}
									\left(
										\frac{1}{\lfloor n^{1+\delta}\rfloor^{K-1}}
										- \frac{\left(\tfrac{t_0}{T}\right)^{4H-K+1}}{\lfloor n^{1+\delta}\rfloor^{K-1}}
									\right)
							\right].
		\ea $$
		This converges strictly faster than $\tfrac{1}{n}$ for all $K \geq 2$. Note that if $K=2$ and $\delta=0$, it only converges at a rate of $\tfrac{1}{n}$. Thus we sample at a rate strictly faster than $\tfrac{1}{n}$. 
		Next we generalize the $i<j$ terms:
		$$\ba
			&		\mathbb{E}[M_{t_{i-1}}^{\alpha_1}M_{t_{j-1}}^{\alpha_2}\Delta M_{t_i}^{\alpha_3}\Delta M_{t_j}^{\alpha_4}] \\
			&=	\mathbb{E}[M_{t_{i-1}}^{\alpha_1}
					((M_{t_{j-1}}-M_{t_i})+M_{t_i})^{\alpha_2}
					\Delta M_{t_i}^{\alpha_3}
					\Delta M_{t_j}^{\alpha_4}] \\
			&=	\mathbb{E}[M_{t_{i-1}}^{\alpha_1}
					((M_{t_{j-1}}-M_{t_i})+\Delta M_{t_i}+M_{t_{i-1}})^{\alpha_2}
					\Delta M_{t_i}^{\alpha_3}
					\Delta M_{t_j}^{\alpha_4}] \\
		\ea$$
		Now we need cases:
		\be
			\item If $\alpha_2=2$ then $\alpha_4=0$ and
				$$\ba
					&		\mathbb{E}[M_{t_{i-1}}^{\alpha_1}
							((M_{t_{j-1}}-M_{t_i})+\Delta M_{t_i}+M_{t_{i-1}})^2
							\Delta M_{t_i}^{\alpha_3}] \\
					=&	\mathbb{E}[M_{t_{i-1}}^{\alpha_1}\Delta M_{t_i}^{\alpha_3}
							((M_{t_{j-1}}-M_{t_i})^2+\Delta M_{t_i}^2+M_{t_{i-1}}^2 \\
								&+2(M_{t_{j-1}}-M_{t_i})\Delta M_{t_i}+2(M_{t_{j-1}}-M_{t_i})M_{t_{i-1}}+2\Delta M_{t_i}M_{t_{i-1}})]\\
					=&	\mathbb{E}[M_{t_{i-1}}^{\alpha_1}\Delta M_{t_i}^{\alpha_3}(M_{t_{j-1}}-M_{t_i})^2]
								+\mathbb{E}[M_{t_{i-1}}^{\alpha_1}\Delta M_{t_i}^{\alpha_3+2}]\\
								&+\mathbb{E}[M_{t_{i-1}}^{\alpha_1+2}\Delta M_{t_i}^{\alpha_3}] 
								+2\mathbb{E}[M_{t_{i-1}}^{\alpha_1}\Delta M_{t_i}^{\alpha_3+1}(M_{t_{j-1}}-M_{t_i})]\\
								&+2\mathbb{E}[M_{t_{i-1}}^{\alpha_1+1}\Delta M_{t_i}^{\alpha_3}(M_{t_{j-1}}-M_{t_i})]
								+2\mathbb{E}[M_{t_{i-1}}^{\alpha_1+1}\Delta M_{t_i}^{\alpha_3+1}].\\
				\ea$$
				Using the independence of disjoint increments of $(M_t)$ and then that $(M_t)$ is centered, this is
				$$\ba
					&		\mathbb{E}[M_{t_{i-1}}^{\alpha_1}]\mathbb{E}[\Delta M_{t_i}^{\alpha_3}]\mathbb{E}[(M_{t_{j-1}}-M_{t_i})^2]
								+\mathbb{E}[M_{t_{i-1}}^{\alpha_1}]\mathbb{E}[\Delta M_{t_i}^{\alpha_3+2}]\\
								&+\mathbb{E}[M_{t_{i-1}}^{\alpha_1+2}]\mathbb{E}[\Delta M_{t_i}^{\alpha_3}] 
								+2\mathbb{E}[M_{t_{i-1}}^{\alpha_1}]\mathbb{E}[\Delta M_{t_i}^{\alpha_3+1}]\mathbb{E}[M_{t_{j-1}}-M_{t_i}]\\
								&+2\mathbb{E}[M_{t_{i-1}}^{\alpha_1+1}]\mathbb{E}[\Delta M_{t_i}^{\alpha_3}]\mathbb{E}[M_{t_{j-1}}-M_{t_i}]
								+2\mathbb{E}[M_{t_{i-1}}^{\alpha_1+1}]\mathbb{E}[\Delta M_{t_i}^{\alpha_3+1}]\\
					=&	\mathbb{E}[M_{t_{i-1}}^{\alpha_1}]\mathbb{E}[\Delta M_{t_i}^{\alpha_3}]c_M(t_{j-1}^{2-2H}-t_i^{2-2H})
								+\mathbb{E}[M_{t_{i-1}}^{\alpha_1}]\mathbb{E}[\Delta M_{t_i}^{\alpha_3+2}]\\
								&+\mathbb{E}[M_{t_{i-1}}^{\alpha_1+2}]\mathbb{E}[\Delta M_{t_i}^{\alpha_3}] 
								+2\mathbb{E}[M_{t_{i-1}}^{\alpha_1+1}]\mathbb{E}[\Delta M_{t_i}^{\alpha_3+1}].
				\ea$$
				If $\alpha_1=\alpha_3=1$ then we have
				$$\ba
					&		\mathbb{E}[M_{t_{i-1}}^{\alpha_1}]\mathbb{E}[\Delta M_{t_i}^{\alpha_3}]c_M(t_{j-1}^{2-2H}-t_i^{2-2H})
								+\mathbb{E}[M_{t_{i-1}}^{\alpha_1}]\mathbb{E}[\Delta M_{t_i}^{\alpha_3+2}]\\
								&+\mathbb{E}[M_{t_{i-1}}^{\alpha_1+2}]\mathbb{E}[\Delta M_{t_i}^{\alpha_3}] 
								+2\mathbb{E}[M_{t_{i-1}}^{\alpha_1+1}]\mathbb{E}[\Delta M_{t_i}^{\alpha_3+1}] \\
					=&	2\mathbb{E}[M_{t_{i-1}}^2]\mathbb{E}[\Delta M_{t_i}^2] \\
					\approx&	
							2c_M^2t_i^{2-2H}(2-2H)t_i^{1-2H}\Delta t \\
					=&	2c_M^2(2-2H)t_i^{3-4H}\Delta t. \\
				\ea$$
				If $\alpha_1=0$ then $\alpha_3=2$ and then
				$$\ba
					&		\mathbb{E}[M_{t_{i-1}}^{\alpha_1}]\mathbb{E}[\Delta M_{t_i}^{\alpha_3}]c_M(t_{j-1}^{2-2H}-t_i^{2-2H})
								+\mathbb{E}[M_{t_{i-1}}^{\alpha_1}]\mathbb{E}[\Delta M_{t_i}^{\alpha_3+2}]\\
								&+\mathbb{E}[M_{t_{i-1}}^{\alpha_1+2}]\mathbb{E}[\Delta M_{t_i}^{\alpha_3}] 
								+2\mathbb{E}[M_{t_{i-1}}^{\alpha_1+1}]\mathbb{E}[\Delta M_{t_i}^{\alpha_3+1}] \\
					\approx&	c_M^2[(2-2H)t_i^{1-2H}\Delta t (t_j^{2-2H}-t_i^{2-2H})
								+3(2-2H)^2t_i^{2-4H}\Delta t^2 \\
								&+(2-2H)t_i^{3-4H}\Delta t]\\
					=&		c_M^2[(2-2H)t_i^{1-2H} t_j^{2-2H}\Delta t
								+3(2-2H)^2t_i^{2-4H}\Delta t^2].
				\ea$$
				Finally, if $\alpha_1=2$ and $\alpha_3=0$ then
				$$\ba
					&		\mathbb{E}[M_{t_{i-1}}^{\alpha_1}]\mathbb{E}[\Delta M_{t_i}^{\alpha_3}]c_M(t_{j-1}^{2-2H}-t_i^{2-2H})
								+\mathbb{E}[M_{t_{i-1}}^{\alpha_1}]\mathbb{E}[\Delta M_{t_i}^{\alpha_3+2}]\\
								&+\mathbb{E}[M_{t_{i-1}}^{\alpha_1+2}]\mathbb{E}[\Delta M_{t_i}^{\alpha_3}] 
								+2\mathbb{E}[M_{t_{i-1}}^{\alpha_1+1}]\mathbb{E}[\Delta M_{t_i}^{\alpha_3+1}] \\
					\approx&	c_M^2[t_i^{2-2H}(t_j^{2-2H}-t_i^{2-2H})
								+t_i^{2-2H}(2-2H)t_i^{1-2H}\Delta t
								+3t_i^{4-4H}] \\
					=&		c_M^2[t_i^{2-2H}t_j^{2-2H}
								+(2-2H)t_i^{3-4H}\Delta t
								+2t_i^{4-4H}]. \\
				\ea$$
			\item If $\alpha_2=1$ then
				$$\ba
					&		\mathbb{E}[M_{t_{i-1}}^{\alpha_1}
							((M_{t_{j-1}}-M_{t_i})+\Delta M_{t_i}+M_{t_{i-1}})
							\Delta M_{t_i}^{\alpha_3}\Delta M_{t_j}^{\alpha_4}] \\
					=&	\mathbb{E}[
							M_{t_{i-1}}^{\alpha_1}\Delta M_{t_i}^{\alpha_3}\Delta M_{t_j}^{\alpha_4}(M_{t_{j-1}}-M_{t_i})
							+M_{t_{i-1}}^{\alpha_1}\Delta M_{t_i}^{\alpha_3}\Delta M_{t_j}^{\alpha_4}\Delta M_{t_i} \\
							&+M_{t_{i-1}}^{\alpha_1}\Delta M_{t_i}^{\alpha_3}\Delta M_{t_j}^{\alpha_4}M_{t_{i-1}}] \\
					=&	\mathbb{E}[M_{t_{i-1}}^{\alpha_1}]\mathbb{E}[\Delta M_{t_i}^{\alpha_3}]
								\mathbb{E}[M_{t_{j-1}}-M_{t_i}]\mathbb{E}[\Delta M_{t_j}^{\alpha_4}] \\
							&+\mathbb{E}[M_{t_{i-1}}^{\alpha_1}]\mathbb{E}[\Delta M_{t_i}^{\alpha_3+1}]\mathbb{E}[\Delta M_{t_j}^{\alpha_4}]
							+\mathbb{E}[M_{t_{i-1}}^{\alpha_1+1}]\mathbb{E}[\Delta M_{t_i}^{\alpha_3}]\mathbb{E}[\Delta M_{t_j}^{\alpha_4}] \\
					=&	\mathbb{E}[M_{t_{i-1}}^{\alpha_1}]\mathbb{E}[\Delta M_{t_i}^{\alpha_3+1}]\mathbb{E}[\Delta M_{t_j}^{\alpha_4}]
							+\mathbb{E}[M_{t_{i-1}}^{\alpha_1+1}]\mathbb{E}[\Delta M_{t_i}^{\alpha_3}]\mathbb{E}[\Delta M_{t_j}^{\alpha_4}]. \\
				\ea$$
				If $\alpha_1=0$ then $\alpha_3=1$ and $\alpha_4=2$ and we have
				$$
					\mathbb{E}[\Delta M_{t_i}^2]\mathbb{E}[\Delta M_{t_j}^2]
							+\mathbb{E}[M_{t_{i-1}}]\mathbb{E}[\Delta M_{t_i}]\mathbb{E}[\Delta M_{t_j}^2]
					\approx	
					c_M^2(2-2H)^2t_i^{1-2H}t_j^{1-2H}\Delta t^2.
				$$
				If $\alpha_1=1$ then $\alpha_3=\alpha_4=1$ and we have
				$$
							\mathbb{E}[M_{t_{i-1}}]\mathbb{E}[\Delta M_{t_i}^2]\mathbb{E}[\Delta M_{t_j}]
							+\mathbb{E}[M_{t_{i-1}}^2]\mathbb{E}[\Delta M_{t_i}]\mathbb{E}[\Delta M_{t_j}] = 0.
				$$
				Finally, if $\alpha_1=2$ then $\alpha_3=0$ and $\alpha_4=1$ and we have
				$$
							\mathbb{E}[M_{t_{i-1}}^2]\mathbb{E}[\Delta M_{t_i}]\mathbb{E}[\Delta M_{t_j}]
							+\mathbb{E}[M_{t_{i-1}}^3]\mathbb{E}[\Delta M_{t_j}] = 0.
				$$
			\item If $\alpha_2=0$ then $\alpha_4=2$ and we have
				$$
							\mathbb{E}[M_{t_{i-1}}^{\alpha_1}
							\Delta M_{t_i}^{\alpha_3}
							\Delta M_{t_j}^2]
							=
							\mathbb{E}[M_{t_{i-1}}^{\alpha_1}]\mathbb{E}[\Delta M_{t_i}^{\alpha_3}]\mathbb{E}[\Delta M_{t_j}^2]
				$$
				and so if $\alpha_1=0$ then $\alpha_3=2$ and we have
				$$
					\mathbb{E}[\Delta M_{t_i}^2]\mathbb{E}[\Delta M_{t_j}^2]
					\approx	
					c_M^2(2-2H)^2t_i^{1-2H}t_j^{1-2H}\Delta t^2.
				$$
				If $\alpha_1=1$ then $\alpha_3=1$ and we have
				$$
					\mathbb{E}[M_{t_{i-1}}]\mathbb{E}[\Delta M_{t_i}]\mathbb{E}[\Delta M_{t_j}^2]=0.
				$$
				Finally, if $\alpha_1=2$ then $\alpha_3=0$ and
				$$
					\mathbb{E}[M_{t_{i-1}}^2]\mathbb{E}[\Delta M_{t_j}^2]
					\approx 
					c_M^2(2-2H)t_i^{2-2H}t_j^{1-2H}\Delta t.
				$$
		\ee
		After incorporating the $\Psi_{t}$ terms, one term emerges to cancel with the term
		$$ C^4\sum\sum t_i^{2H-1}t_j^{2H-1}(\Delta t)^2 $$
		in \eqref{lasttwoterms}. 
		Otherwise, all remaining terms are of the form $$\sum\sum t_i^{2H-M}t_j^{2H-N}(\Delta t)^{M+N},$$ for combinations of $N,M\in\{1,2\}$ except $M=N=1$.
		For terms of this form, setting
		$t_i=\tfrac{iT}{m}$ and
		$t_j=\tfrac{jT}{m}$, we have
		$$
		\ba
			&  \sum\limits_{i=i_0}^m \sum\limits_{j=i+1}^m t_i^{2H-M} t_j^{2H-N} (\Delta t)^{M+N} \\
			=& \left(\tfrac{T}{m}\right)^{4H} \sum\limits_{i=i_0}^m i^{2H-M} \sum\limits_{j=i+1}^m j^{2H-N} \\
			\leq& \left(\tfrac{T}{m}\right)^{4H} 
										\sum\limits_{i=i_0}^m i^{2H-M}  
										\int_i^m \! x^{2H-N} \, \mathrm{d}x. \\
			=& \tfrac{T^{4H}}{2H-N+1} 
										\left[
											\tfrac{1}{m^{2H+N-1}} \sum\limits_{i=i_0}^m i^{2H-M} 
											- \tfrac{1}{m^{4H}} \sum\limits_{i=i_0}^m i^{4H-N-M+1} 
										\right] \\
			\approx& \tfrac{T^{4H}}{2H-N+1} 
										\left[
											\tfrac{1}{m^{2H+N-1}} 
												\left[i_0^{2H-M} + \int_{i_0}^m \! x^{2H-M} \, \mathrm{d}x \right] \right.\\
											&-\left. \tfrac{1}{m^{4H}} 
												\left[i_0^{4H-M-N+1} + \int_{i_0}^m \! x^{4H-N-M+1} \, \mathrm{d}x \right] 
										\right].\\
			=& \tfrac{T^{4H}}{2H-N+1} 
						\Bigg[ 
							\frac{\left(\tfrac{t_0}{T}\right)^{2H-M}}{m^{M+N-1}}
							+ \tfrac{1}{2H-M+1}
									\left(
										\frac{1}{m^{M+N-2}}
										- \frac{\left(\tfrac{t_0}{T}\right)^{2H-M+1}}{m^{M+N-2}}
									\right) \\
							& - \frac{\left(\tfrac{t_0}{T}\right)^{4H-M-N+1}}{m^{M+N-1}}
							- \tfrac{1}{4H-M-N+2}
									\left(
										\frac{1}{m^{M+N-2}}
										- \frac{\left(\tfrac{t_0}{T}\right)^{4H-M-N+2}}{m^{M+N-2}}
									\right)
						\Bigg].
		\ea
		$$

		This converges strictly faster than $\tfrac{1}{n}$ for all $N,M\in\{1,2\}$, excluding $M=N=1$, as required.
		Therefore we have proven that
		$$\left|\left|\sum\limits_{i=i_0}^m (\Delta V_{t_i})^2 -I^*\right|\right|_2^2$$
		as given in \eqref{allthreeterms}, is strictly summable and therefore by Chebyshev's inequality, for any $\epsilon>0$,
		$$\sum_{n=1}^\infty P\left(\left|\sum_{i=1}^n \left(\Delta V_{t_i}\right)^2-I^*\right|>\epsilon\right)
			\leq \sum_{n=1}^\infty \frac{1}{\epsilon^2} \left|\left|\sum\limits_{i=i_0}^m (\Delta V_{t_i})^2 -I^*\right|\right|_2^2
			< \infty.$$
		Finally, by the Borel-Cantelli Lemma,
		$$\lim_{n\rightarrow\infty}\sum_{i=1}^n \left(\Delta V_{t_i}\right)^2 = I^*$$
		almost surely.
	\eprof
	\label{sec:appendix}

	\printbibliography

\end{document}